\documentclass[a4paper,dvips,11pt]{article}
\usepackage{epsf}
\usepackage[dvips]{graphicx}
\usepackage{tabularx}
\usepackage{latexsym}
\usepackage{amsmath,amssymb,exscale}
\usepackage{array,multicol}
\numberwithin{equation}{section}

\def\ov{\overline}

\def\dalemb#1#2{{\vbox{\hrule height .#2pt
         \hbox{\vrule width.#2pt height#1pt \kern#1pt
                 \vrule width.#2pt}
         \hrule height.#2pt}}}

\let\a=\alpha    
   \let\i=\iota 
    \let\p=\pi

\let\F=\Phi

 \def\bd{\begin{document}} \def\ed{\end{document}}
\def\ds{\documentstyle} \let\fr=\frac \let\bl=\bigl \let\br=\bigr
\let\Br=\Bigr \let\Bl=\Bigl
\let\bm=\bibitem
\let\na=\nabla
\let\pa=\partial
\let\ov=\overline
\def\ie{{\it i.e.\ }}
\def\tr{{\mbox{\rm tr}}}
\newcommand{\be}{\begin{equation}}
\newcommand{\ee}{\end{equation}}
\newcommand{\beba}{\begin{equation}\begin{array}{lcl}}
\newcommand{\eaee}{\end{array}\end{equation}}
\newcommand{\bea}{\begin{eqnarray}}
\newcommand{\eea}{\end{eqnarray}}
\newcommand{\ba}{\begin{array}}
\newcommand{\ea}{\end{array}}
\newcommand{\td}{\tilde}
\newcommand{\norsl}{\normalsize\sl}
\newcommand{\ns}{\normalsize}
\newcommand{\refs}[1]{(\ref{#1})}
\def\simlt{\mathrel{\lower2.5pt\vbox{\lineskip=0pt\baselineskip=0pt
            \hbox{$<$}\hbox{$\sim$}}}}
\def\simgt{\mathrel{\lower2.5pt\vbox{\lineskip=0pt\baselineskip=0pt
            \hbox{$>$}\hbox{$\sim$}}}}
\def\A{{\cal A}}
\def\a{{\mathcal a}}
\def\V{{\cal V}}
\def\F{{\cal F}}
\def\p{{\mathcal \phi}}
\def\L{{\mathcal L}}
\def\M{{\mathcal M}}
\def\bD{{\ov {\rm D}}}
\def\bO{{\ov {\rm O}}}
\def\bOp{{\ov {\rm O'}}}
\def\O{{ {\rm O}}}

\title{
\vspace*{-0.8cm}
\begin{flushright}
\normalsize{CERN-PH-TH/2004-240\\ 
\texttt{hep-th/0412008}}\\
\end{flushright}
\vspace{1cm}
\Large\textbf{Moduli stabilization from magnetic fluxes in type~I 
string theory}
\author{\large
{\bf I.~Antoniadis~$^1$\footnote{On leave of absence from CPHT,
Ecole Polytechnique, UMR du CNRS 7644.}, T. Maillard$^{1,2}$}\\ \\
\emph{$^1$CERN Theory Division
   CH--1211, Gen{\`e}ve 23, Switzerland }\\
\emph{$^2$Institut f\"ur Theoretische Physik, ETH H\"onggerberg}\\
\emph{ CH--8093\, Z\"urich, Switzerland}}}
\date{}
\begin{document}
\maketitle
\thispagestyle{empty}
\vspace*{.5cm}

\begin{abstract}
We show that type I string theory 
compactified in four dimensions in the presence of constant internal 
magnetic fields possesses ${\cal N}=1$ supersymmetric vacua, in which 
all K\"ahler class and complex structure closed string moduli are 
fixed. Furthermore, their values can be made arbitrarily large by
a suitable tuning of the quantized magnetic fluxes.
We present an explicit example  for the toroidal 
compactification on $T^6$ and discuss Calabi-Yau generalizations. 
This mechanism can be complementary to other stabilization methods 
using closed string fluxes but has the advantage of having an exact 
string description and thus a validity away from the low-energy 
supergravity approximation. Moreover, it  can be easily implemented 
in constructions of string models based on intersecting $D$-branes.
\end{abstract}
\date

\newpage

\section{Introduction}

The problem of moduli stabilization in 
string compactifications is of great interest and significant 
progress has been made recently based on closed string 
fluxes~\cite{Kachru:2002he,Review}. For instance, by a suitable 
choice of NS-NS (Neveu Schwarz - Neveu Schwarz) and 
R-R (Ramond - Ramond) 3-form fluxes, one 
can find ${\cal N}=1$ supersymmetric vacua where all complex 
structure moduli, as well as the dilaton, are 
fixed~\cite{Kachru:2002he}. A 
disadvantage of the method is that there is no exact string 
description of such fluxes and thus the analysis is restricted to the 
lowest order in $\alpha'$ expansion, described by the effective field 
theory. Moreover, generalization of the stabilization mechanism to 
K\"ahler class moduli requires introduction of non-perturbative 
effects which are again treated in the low-energy supergravity 
approximation~\cite{Kachru:2003aw}.

In this work, we present an alternative mechanism of moduli 
stabilization based on open string constant magnetic 
backgrounds that have an exact description in string 
theory~\cite{Fradkin:1985qd,Bachas:1995ik}. 
In fact, magnetic fluxes can be turned on in any 2-cycle of the 
internal compactification manifold. In the simplest case, 
magnetic backgrounds on (1,1)-cycles fix K\"ahler class 
moduli~\cite{Blumenhagen:2003vr}, while backgrounds on 
holomorphic (2,0)-cycles fix complex structure moduli. 
This is illustrated in a simple example of a toroidal 
compactification on $T^6$, with a suitable choice of internal 
magnetic fields that must be turned on in several Cartan 
directions of the higher-dimensional ($D9$-brane) gauge group 
(such as $SO(32)$). In this example,
we demonstrate that all closed string moduli, but the 
dilaton, can be fixed in a vacuum preserving ${\cal N}=1$ 
supersymmetry, in terms of the magnetic fluxes.

There are three important ingredients in this stabilization mechanism.
The first is the non linearity of the Dirac-Born-Infeld (DBI) action
which allows to fix all K\"ahler class moduli, including the overall
volume. The second is the fact that magnetized $D9$-branes in
a six dimensional internal manifold induce partly negative
five-brane tension, which makes possible tadpole cancellation in a
supersymmetric configuration, without adding lower dimensional 
$D$-branes or orientifold planes (such as $D5$ or $O5$).
The third is the use of magnetic fluxes in non-diagonal
directions of the internal space which allows to fix all non-diagonal
components of the metric. For instance, in the particular
example we present in this work, the metric is fixed to a
completely diagonal form.
It is also interesting to point out that
despite the use of $\alpha'$ corrections (non-linear terms), it
turns out that the moduli can be stabilized at arbitrarily large
values by tuning appropriately the quantized magnetic fluxes.

This mechanism can be obviously combined with the presence of
closed type IIB string 3-form fluxes, allowing to fix the dilaton and the 
complex structure of more general compactification manifolds. 
It can also be easily implemented in
constructions of string models based on $D$-branes at angles.
Finally, it should be a priori possible to generalize it in order
to stabilize moduli in vacua with broken supersymmetry, 
following arguments similar to those of 
refs.~\cite{Kachru:2003aw,Kachru:2002he} for 3-form fluxes.

Our paper is organized as follows. 
In Section \ref{notation}, we recall the main properties of the 
six-dimensional toroidal compactification and its moduli 
space~\cite{Moore:1998pn,Candelas:1990pi}. 
In Section \ref{sec:fluxes}, we consider open string propagation 
in the presence of constant internal magnetic 
fields~\cite{Fradkin:1985qd} 
and summarize the conditions for unbroken supersymmetry. In 
Section \ref{sec:susy_vac}, we extract the equations that guarantee 
the existence of one unbroken supersymmetry preserved by a stack of 
magnetized D9-branes~\cite{Marino:1999af,Angelantonj:2000hi} and 
study the conditions for which this supersymmetry is the same with 
the one preserved by lower dimensional $D$-branes. 
In Section \ref{sec:moduli_stab}, we analyze the above 
conditions and show how the moduli can be stabilized 
by a suitable choice of several magnetic backgrounds in 
the Cartan subalgebra of the gauge group. We also study
the tadpole cancellation conditions which are required for 
consistency of type I string vacua.
In Section \ref{sec:example}, 
we come back to the toroidal $T^6$ compactification and present an 
explicit solution of the supersymmetry equations and tadpole 
cancellation conditions that fixes all K\"ahler class and complex 
structure moduli in terms of the magnetic fields.
For completeness, we also give an explicit numerical solution
in Appendix A. In Section \ref{sec:led}, we show that the
above solution can be appropriately ``rescaled" to generate
arbitrarily large values for the internal radii. Moreover,
in Appendix B, we present a corresponding numerical example.
In Section \ref{sec:RR}, we describe the stabilization of the 
R-R moduli. In particular, we show that those complexifying 
the K\"ahler class metric deformations are stabilized by 
being absorbed in the (anomalous) magnetized $U(1)$ 
gauge fields that become massive~\cite{Angelantonj:2000hi}. 
Finally in Section \ref{sec:CY}, we discuss generalizations of the 
method to orbifolds and more general compactifications, the 
problem of fixing the dilaton and other concluding remarks.

\section{Torus compactification}\label{notation}
\subsection{Parametrization of $T^6$}

Consider a six-dimensional torus $T^{6}$ having six coordinates 
$x^i$, 
$y_i$ with $i=1,2, 3$ and periodicity normalized to unity  
$x^i=x^i+1$, $y_i=y_i+1$~\cite{Moore:1998pn}. 
We choose then the orientation 
\be
\int_{T^6} dx^1\wedge dy_1\wedge dx^2\wedge dy_2\wedge dx^3\wedge 
dy_3  = 1
\label{orientation}
\ee
and define the basis of the cohomology $H^3(T^6,\mathbb{Z})$
\bea
\alpha_0 & = &  dx^1\wedge dx^2\wedge dx^3 \nonumber \\
\alpha_{ij} & = & \frac{1}{2}\epsilon_{ilm}dx^l\wedge dx^m\wedge dy_j 
\label{H3basis} \\
\beta^{ij} & = & -\frac{1}{2}\epsilon^{ilm}dy_l\wedge dy_m\wedge dx^j 
\nonumber\\
\beta^0 & = & dy_1\wedge dy_2\wedge dy_3, \nonumber
\eea
forming a symplectic structure on $T^6$:
\be
\int_{T^6} \alpha_a \wedge \beta^b = -\delta_a^b\, \, ,
\,\,\,\, \textrm{for}\,\,  a,b =1,\cdots,h_3/2\, , 
\label{symplectic_structure}
\ee
with $h_3 = 20$, the dimension of the cohomology 
$H^3(T^6,\mathbb{Z})$. 

We can also choose complex coordinates 
\be z^i = x^i + \tau^{ij}y_j,
\label{complex_structure}
\ee 
where $\tau^{ij}$ is a complex $3 \times 3$ matrix 
parametrizing the complex structure. In  this basis, the cohomology 
$H^3(T^6,\mathbb{Z})$ decomposes in four different cohomologies 
corresponding to the purely holomorphic parts and those with mixed 
indices:
\be
H^3(T^6,\mathbb{Z}) = H^{3,0}(T^6,\mathbb{Z})\oplus 
H^{2,1}(T^6,\mathbb{Z})\oplus H^{1,2}(T^6,\mathbb{Z})\oplus 
H^{0,3}(T^6,\mathbb{Z}).
\ee
The purely holomorphic cohomology $H^{3,0}$ is one-dimensional and is 
formed by the holomorphic three-form $\Omega$ for which we choose the 
normalization
\be
\Omega = dz^1\wedge dz^2 \wedge dz^3. \label{omega}
\ee

\subsection{Moduli space}

Consider now  the deformations  $\delta g_{ij}$,  $\delta 
g_{i\bar{j}}$ of the flat metric on $T^{6}$~\cite{Candelas:1990pi} .
The purely holomorphic variation $\delta g_{ij}$ is parametrized by 
elements of the cohomology $H^{2,1}$. 
Indeed, one can represent the corresponding $(2,1)$ deformations as: 
\be
\chi_{{\tilde a} \,\kappa\lambda \bar{\mu}} \equiv 
-\frac{1}{2} \Omega_{\,\,\,\,\kappa\lambda}^{\bar{\nu}}\, \frac{\partial 
g_{\bar{\mu}\bar{\nu}}}{\partial \tau^{\tilde a}}\, ,\quad  
{\tilde a} = 1,\ldots,h_{2,1}, 
\label{chi_deformation}
\ee
where $h_{2,1} = {\rm dim}\, H^{2,1}$.

Next, we  write  the space of complex structures in terms of the 
periods of the holomorphic 3-form. Define the canonical homology 
basis $(A_{a},B^{b})$ dual to the basis of $H^3$ given in 
(\ref{H3basis}) by  rewriting the symplectic structure 
(\ref{symplectic_structure}) as
\bea
\int_{A_{a}}\alpha_{b} &=:&\int_{T^6}\alpha_{b}\wedge \beta^a =  
-\delta^a_{b}\nonumber \\
\int_{B^{a}}\beta^{b} &=:&\int_{T^6}\beta^{b}\wedge \alpha_a =  
\delta_{a}^{b}\nonumber
\eea
The periods of the holomorphic 3-form are then defined as
\be
\tau^{a} = \int_{A_{a}}\Omega\, .
\ee 
Comparing with the definition (\ref{omega}) for $\Omega$, we can
identify  the matrix elements of $\tau$ in the 
parametrization (\ref{complex_structure}) of the complex basis with 
the 
periods of $\Omega$. The dimension of the space of complex structure 
moduli is then given by the dimension of the cohomology $H^{2,1}$ on 
the torus $T^{6}$, 
$h_{2,1}=9$.

On the other hand, the metric variation  $\delta g_{i\bar{j}}$
 is parametrized by elements of the $H^{1,1}$ cohomology
\be
J = i\delta g_{i \bar{j}} dz^{i} \wedge dz^{\bar{j}}.
\ee
Let $e_{A}$, $A=1,\ldots,h_{1,1}$, be a basis of $H^{1,1}$. In our 
case, one 
can choose:
\be
e^{i\bar{j}} = idz^{i} \wedge dz^{\bar{j}}\, ,
\label{e_basis}
\ee
so that the K\"ahler form can be parametrized as
\be
J = J_{i\bar{j}}e^{i\bar{j}}\, .
\label{kaehler_structure}
\ee
The elements $J_{i\bar{j}}$ satisfy the reality condition
$J_{i\bar{j}}^\dagger=J_{j\bar{\i}}$, implying that $J$ depends
on nine real parameters. They can be used to parametrize the
space of  K\"ahler deformations whose dimension is given 
by the dimension of the cohomology $H^{1,1}$ on the torus $T^{6}$,  
$ h_{1,1} = 9$.

\section{Type I string theory: Fluxes and conserved supercharges}
\label{sec:fluxes}


\subsection{Magnetic fluxes}\label{subsec:gauge_field}

Consider a stack of $N$ coincident $D9$ branes, giving rise to 
a $U(N)$ $\mathcal{N}=4$ supersymmetric gauge theory. Pick up a 
$U(1)$ subgroup  in 
the Cartan subalgebra of $U(N)$ with gauge potential $A$, and
turn on a constant magnetic field. Thus, the 
corresponding field strength $F_{kl}$ is constant and  $A_k = 
-\frac{1}{2}F_{kl}u^l$, where $u^l$ stands for all six coordinates of 
$T^6$, $x^i$ and $y^i$. 
This constant magnetic background couples to the boundary of the 
open string on the brane by quadratic terms in the world-sheet 
action $S_{ws}$~\cite{Fradkin:1985qd}. The corresponding 
conformal field theory can therefore be solved exactly:
\bea
S_{ws} &=& -\frac{1}{4\pi\alpha'}\int_\Sigma dt 
d\sigma\left(\partial_\alpha X^\mu 
\partial^\alpha X_\mu - 
i\bar{\psi^\mu}\rho^\alpha\partial_\alpha\psi_\mu\right)
\nonumber \\
     && -\int dtq_L F_{kl}\left(X^k\partial_t 
X^l-\frac{i}{2}\bar{\psi^k}\rho^0\psi^l \right)_{\sigma=0} 
\label{worldsheet_action} \\ 
     && -\int dtq_R F_{kl}\left(X^k\partial_t 
X^l-\frac{i}{2}\bar{\psi^k}\rho^0\psi^l\right)_{\sigma=\pi} \nonumber
\eea
where $\alpha '$ is the Regge slope, $\psi^{\mu}$ are the real 
Majorana fermionic superpartners of the coordinates $X^{\mu}$ 
and  $\rho^{\alpha}$ with $\alpha = 0,1$
are the two-dimensional gamma-matrices. 
The indices $k,l$ run over the magnetized dimensions
$k,l = 4,\cdots , 9$, whereas the indices $\mu,\nu$ run  over all 
ten-dimensional spacetime coordinates $\mu,\nu = 0,\cdots,9$. 
The couplings of the left and right endpoints of the open string to 
the background are given by the corresponding charges $q_L$ and $q_R$.

The field $F_{kl}$ corresponds to a non trivial $U(1)$ gauge bundle 
over the torus $T^6$ with transition function around the cycles 
$u_k$:
\be
A_k{\big |}_{u^l+1} = \left(A_k-ie^{-iq\theta} 
\partial_ke^{iq\theta}\right){\bigg |}_{u^l}\, , \,\,\,\,\, 
\theta = 
F_{kl}u^l
\ee
with $q = q_L+q_R$. Imposing the phase  over 
each cycle $u^k$ to be single-valued leads to the usual Dirac 
quantization condition
\be
q\cdot F_{kl} = 2\pi m_{kl},\,\,\,\, \forall\, k,l = 4, 
\ldots , 9\, ,
\label{dirac1}
\ee
where $m_{kl}$ are integers corresponding to the first Chern 
class of the $U(1)$ gauge bundle. We can extend this quantization 
condition
by taking into account the multiplicity $n_{kl}$ of the winding 
around 
the 2-cycles $[u^k,u^l]$. Equation (\ref{dirac1}) is then modified 
as~\cite{Blumenhagen:2000vk}:
\be
q\cdot F_{kl} = 2\pi \frac{m_{kl}}{n_{kl}},\quad
\forall\, k,l = 4, \ldots , 9\, .
\label{dirac2}
\ee
Finally, we can easily extend our analysis in the presence
of a NS-NS $B$-field, which in type I compactifications has quantized
values~\cite{Bquant}. Its effect consists of replacing 
$q\cdot F$ by $q\cdot F- B/(2\pi\alpha')$. This amounts to shifting
the integers $m_{kl}$ in the above quantization condition by 
$m_{kl}-bn_{kl}$, where $b$ is quantized and equals $1/2$ or $0$ 
in the toroidal case [see eq.~(\ref{FB}) below].


\subsection{Supersymmetry}

The presence of constant internal magnetic fields generically 
breaks  supersymmetry by shifting the masses of the four 
dimensional scalars and fermions~\cite{Bachas:1995ik}. 
{}For suitable choice of the fluxes and moduli
however, a four-dimensional supersymmetric theory can be 
recovered~\cite{Angelantonj:2000hi}. In this section, we analyze the 
conditions under which a supersymmetric vacuum can exist.

Consider the supersymmetric $D9$ brane action of type IIB 
string theory~\cite{Bergshoeff:1996tu}:
\be
V = V_{DBI} + V_{WZ}\, ,
\label{action_superspace}
\ee
where 
\be
V_{DBI} = -\int [d^{10}X] 
e^{-\phi}\sqrt{-\det(g_{\mu\nu}+\mathcal{F}_{\mu\nu})} 
\ee
is the Dirac-Born-Infeld (DBI) action extended in the 
superspace and $V_{WZ}$ is the corresponding Wess-Zumino (WZ) term. 
$\phi$ is the string dilaton and $\mathcal{F}$ is the 2-form field strength:
\be
\mathcal{F}=2\pi\alpha'qF-B\, ,
\label{FB}
\ee
with $F$ the usual field strength  of the $U(1)$ gauge field, 
$F=dA$, on the world-volume of the $D$-brane with charge $q$, whereas 
$B$ is the 2-form 
potential of the NS-NS closed string sector. In the superspace 
formalism, the 
superspace forms may be expanded in terms of the basis 1-forms 
$dZ^M$, 
or equivalently, in the inertial frame basis $E^A = dZ^M E_M^A$, with 
$E_M^A $ being 
the supervielbein of type IIB supergravity. The basis $E^A$ 
decomposes under the action of the Lorentz group into a Lorentz 
vector $E^a$ and a Lorentz spinor $E^{\alpha I}$. For type IIB 
superspace, the latter is a pair of Weyl Majorana spinors, $I=1,2$. 
The metric in the DBI action is defined as 
\be
g_{\mu\nu}= E_{\mu}^aE_{\nu}^b \eta_{ab}\, .
\ee

The superspace extension of the Wess Zumino (WZ) action is similar
\be
I_{WZ} = \int Ce^{\mathcal{F}}\, ,
\ee
where $C$ is a formal sum over the $r$-forms superspace potentials 
$C_{r}$:
\be
C = \sum_{r=0}^{10} C_{r}\, ,
\ee
with $r$ even.

The action (\ref{action_superspace}) is invariant under the 
$\kappa$-symmetry transformations:
\bea
\delta_{\kappa}Z^M E_M^a &=& 0 \nonumber\\
\delta_{\kappa}Z^M E_M^{\alpha} & = & 
[\bar{\kappa}(1+\Gamma)]^{\alpha} \label{kappa_transformation}\nonumber\\
\delta_{\kappa}A_{\mu} & = & E_{\mu}^C \delta_{\kappa} E^D B_{DC}
\\
\delta B &=& \frac{1}{2}E^A E^D \delta_{\kappa} E^C H_{CDA} + 
d(E^A\delta_{\kappa} E^C B_{CA})\nonumber\\
\delta_{\kappa}\phi &=& 0\nonumber
\eea
where $\bar\kappa$ is a ten-dimensional Majorana-Weyl spinor parameter,
$A_\mu$ is the gauge potential,
and the 3-form $H_{ABC}$ is the field strength of $B_{BC}$, $H=dB$.
Its non-zero elements are $H_{\alpha\beta c}=
i\sigma_3\left(\Gamma_c{1+\Gamma^{(11)}\over 2}\right)_{\alpha\beta}$, with $\Gamma_c$ the spacetime gamma-matrices. The matrix $\Gamma$ for a $D9$ brane, appearing in the second relation of (\ref{kappa_transformation}), is
\be
\Gamma = \frac{\sqrt{|g|}}{\sqrt{|g+\mathcal{F}|}}\sum_{n=0}^\infty 
\frac{1}{2^n n!} \gamma^{\mu_1\nu_1 \ldots \mu_n \nu_n} 
\mathcal{F}_{\mu_1 
\nu_1}\ldots \mathcal{F}_{\mu_n \nu_n}J_{(9)}^{(n)}
\label{Gamma}
\ee
with
\bea
J_{(9)}^{(n)} & = & (-1)^n \sigma_3^{n+3}i\sigma_2\otimes 
\Gamma^{(11)}\nonumber \\
\Gamma^{(11)} & = & \frac{1}{10! \sqrt{|g|}}\epsilon^{\mu_1\cdots 
\mu_{10}}\gamma_{\mu_1\cdots\mu_{10}}\, ,
\eea
where $\gamma_{i_1 \cdots i_n} = 
\frac{1}{n!}\gamma_{[i_1}\cdots\gamma_{i_n]} $ denotes complete
antisymmetrization of the corresponding gamma matrices,
and the first term $n=0$ in the sum of (\ref{Gamma}) is reduced to 
$J_{(9)}^{(0)} $.
Here, we denote $|g| = \det{g_{\mu\nu}}$, $|g+\mathcal{F}| = 
\det(g_{\mu\nu}+\mathcal{F}_{\mu\nu})$, $\sigma_i , \,\, i=1,2,3$ the usual 
Pauli matrices,
and $\gamma_{\mu} = E_{\mu}^a \Gamma_a$.
Therefore, the fraction of supersymmetry preserved by the $D9$ branes 
is given by the dimension of the space of solutions of 
~\cite{Bergshoeff:1997kr}:
\be
(1- \Gamma ) \epsilon = 0\, ,
\label{susy_condition1}
\ee
where $\epsilon$ is the spinorial spacetime supersymmetry parameter 
$\epsilon = (\epsilon_1,\epsilon_2)^T$.

On the other hand, considering the $D9$ brane as a source in a 
supergravity configuration, a supersymmetric vacuum asks for a second 
condition.
The expectation value of a supersymmetry transformation of the 
gravitino must vanish in the the chosen background
\be
<\delta_{\epsilon}\Psi_\mu> = 0 \>\, ,
\label{covcst}
\ee
which leads to the usual covariantly constant spinor equation
\be
\nabla_\mu \epsilon = 0\, .
\label{susy_condition2}
\ee
The number of conserved supercharges in our magnetic background is 
then given by the dimension of the space of solutions of the 
covariantly
constant spinors, satisfying:
\be
(1- \Gamma ) \epsilon = 0\, .
\label{susy_condition_last}
\ee

\section{Existence of supersymmetric vacuum}\label{sec:susy_vac}

In this section, we analyze the existence of solutions to the 
equations (\ref{susy_condition2}) and (\ref{susy_condition_last}), 
corresponding to vacua with $\mathcal{N}=1$ 
supersymmetry in four dimensions, and discuss their properties.

\subsection{Solution to the supersymmetry condition }

We first decompose the spinor representation of $SO(1,9)$ under 
$SO(1,3)\otimes SO(6)$ by defining the gamma-matrices:
\bea
\hat\gamma_\mu & = & \Gamma_\mu \otimes 1 \,\,\, , \,\,\, \mu = 0,\ldots,3 \\
\hat\gamma_k   & = & \Gamma^{(4)} \otimes \Gamma_k  \,\,\, , \,\,\, k = 
4,\ldots,9,
\eea
of $SO(1,3)$ and $ SO(6)$, respectively. The chirality operators are 
defined as:
\be
\Gamma^{(4)}  =    i \hat\gamma_0\hat\gamma_1 \hat\gamma_2\hat\gamma_3 \,\,\, , \,\,\,
\Gamma^{(6)}   =   i \hat\gamma_4\hat\gamma_5 
\hat\gamma_6\hat\gamma_7\hat\gamma_8\hat\gamma_9\, ,
\label{4dim_decomposition}
\ee
and have both eigenvalues $\pm 1$. Under this decomposition, the 
Majorana-Weyl spinor $\epsilon_I$, with $I=1,2$,
of $SO(1,9)$ can be written as
\be
\epsilon_{I} = \sum_{i=1}^4 (\psi_{i;I} \cdot \eta_{i;I} + 
\psi_{i;I}^\star \cdot \eta_{i;I}^\star)\, ,
\label{4dim_decomposition_2}
\ee
where $\psi$'s and $\eta$'s are, respectively, Weyl spinors of  
$SO(1,3)$ and $SO(6)$ with positive chirality:
\be
\Gamma^{(4)} \psi_{i;I} = \psi_{i;I} \,\,\,\, , \,\,\,\,
\Gamma^{(6)} \eta_{i;I} = \eta_{i;I}\, ,
\ee
while the $\psi^\star$'s and $\eta\star$'s are their complex conjugates
with opposite chiralities. 
For a magnetic field background $\mathcal{F}$ 
confined in the internal $T^6$, the supersymmetry conditions 
(\ref{susy_condition2}) and (\ref{susy_condition_last}) reduce to
\bea
(1-\Gamma )\eta & = & 0 
\label{susy_condition_last22}\\
\nabla \eta & = & 0\, ,  
\label{susy_condition_last2}
\eea
with $\eta = ( \eta_{i;1} ,\eta_{i;2})^T$. 

Consider now the  case  where there are only two covariantly constant 
spinors of $SO(6)$, $\eta $ and $\eta^{\star}$, conjugate to each 
other and with opposite chirality, which satisfy equations 
(\ref{susy_condition_last22}) 
and  (\ref{susy_condition_last2})~\cite{Marino:1999af}. Using the 
complex basis (\ref{e_basis}) for gamma-matrices, the condition 
(\ref{susy_condition_last2}) 
is then equivalent with the existence of a single holomorphic 3-form 
$\Omega$ and a single K\"ahler form $J$ defined by:
\bea
\gamma_{i\bar{j}}\eta & = & iJ_{i\bar{j}}\eta
\nonumber \\
\gamma_{ijk}\eta& = & \Omega_{ijk}\eta^\star\, .
\label{defOJ}
\eea
Furthermore, the field 
strength $\mathcal{F}$ splits in the complex basis in purely 
holomorphic  $\mathcal{F}_{(2,0)}$, $\mathcal{F}_{(0,2)}$ and mixed 
$\mathcal{F}_{(1,1)}$ parts.  Using (\ref{defOJ}), one can then show 
that eq.~(\ref{susy_condition_last22}) is equivalent 
to~\cite{Marino:1999af}:
\bea
(iJ+\mathcal{F})^{3} & = &e^{i\theta} 
\sqrt{|g_6+\mathcal{F}|}\frac{V_6}{\sqrt{|g_6|}} 
\label{kaehler_condition1}\\
\mathcal{F}_{(2,0)}  & = & 0\, ,
\label{complex_structure_condition1}
\eea
where $V_6$ is the volume form of $T^6$ and $g_6$ is its metric. 

Eqs.~(\ref{kaehler_condition1}) and (\ref{complex_structure_condition1})
impose that the magnetized $D9$ branes preserve ${\mathcal N}=1$
supersymmetry in four dimensions. In order to have more 
supersymmetries, extra conditions have to be imposed, 
related to the requirement of existence of additional covariantly 
constant spinors satisfying (\ref{susy_condition_last22}), 
besides the pair $\eta$ and $\eta^{\star}$.

Eq.~(\ref{kaehler_condition1}) can be rewritten in the form:
\be
\tan{\theta}\left(J\wedge J \wedge \mathcal{F} - 
\mathcal{F}\wedge\mathcal{F} \wedge \mathcal{F}\right) = 
J \wedge J \wedge J - J\wedge 
\mathcal{F} \wedge \mathcal{F}\, ,
\label{kaehler_condition2}
\ee
where the wedge product $A^N$ is defined with an implicit 
normalization factor $1/N!$.
Note that only the $(1,1)$-part of $\mathcal{F}$ appears in this 
formula.  Formally, (\ref{kaehler_condition2}) can be also written as
\be
\rm{Im}\left(e^{-i\theta}\Phi\right)=0\, ,
\label{kaehler_condition3}
\ee
with 
\be
\Phi = 
(iJ+\mathcal{F})\wedge(iJ+\mathcal{F})\wedge(iJ+\mathcal{F})\, .
\label{Phi}
\ee
The constant phase $\theta$ remains so far an unconstrained parameter.

\subsection{$\theta$-parameter and conserved supercharges.}

Let us discuss now the set of $\theta$-angles for which 
 the supersymmetry preserved by the magnetized $D9$ 
branes is the same with the one preserved by lower 
dimensional branes or  orientifold planes $Op$ with $p \le 9$. 
The reason is that the presence of such objects is in general 
necessary for tadpole cancellation~\cite{Angelantonj:2002ct},
as we will see in the next section.

Choosing a basis on the torus $T^6$ where the 
K\"ahler form  $J_{i\bar{j}}$ and the background field strength $\mathcal{F}$ 
are simultaneously diagonal $J_{i\bar{j}} = \delta_{i \bar{j}}$ and
$\mathcal{F}_{i\bar{j}} = f_i\cdot \delta_{i\bar{j}}$, 
with $\delta\equiv{\bf 1}_3\otimes i\sigma_2$, 
$J_{i\bar{j}} = J^{\star}_{j\bar{i}}$, 
$\mathcal{F}_{i\bar{j}} = \mathcal{F}^{\star}_{j\bar{i}}$, 
one can express the field strength components $f_i$ in terms of the 
T-dual angles $\varphi_i$ of wrapped $D6$ branes in the $i$-th two-torus:
\be
\tan \varphi_i = f_i\, .
\ee 
The condition (\ref{kaehler_condition2}) for a supersymmetric 
vacuum then reads:
\be
\sum_i \varphi_i = \frac{3\pi}{2}-\theta\, .
\label{varphi_theta}
\ee
Moreover,  the supersymmetry transformation parameter preserved by 
the magnetized $D9$ 
branes has to satisfy
\be
\epsilon_2 = \Gamma^0 \cdots \Gamma^9 \rho(\mathcal{F}) \epsilon_1,
\label{susy_magnbrane}
\ee
where $\rho(\mathcal{F})$ is the spinorial representation of the 
action of 
the magnetic fluxes~\cite{Mihailescu:2000dn} and equals a phase,
$\rho(\mathcal{F})=e^{i\sum_i\varphi_i}$.

Let us examine now the fraction of  supersymmetry preserved by 
orientifold planes. Consider first the case of $O3$ 
planes~\cite{Blumenhagen:2003vr}, defined by the space of 
fixed points of the orientifold action, which combines the 
world-sheet parity with the usual parity in the six internal 
dimensions. The invariant supersymmetry transformation
parameter under the orientifold action satisfies the equation:
\be
\epsilon_2 = \Gamma^4 \cdots \Gamma^9 \epsilon_1\, .
\ee 
This condition, combined with the fraction (\ref{susy_magnbrane}) of 
the supersymmetry 
conserved by the magnetized $D9$ branes gives rise to 
\be
\rho(\mathcal{F})\eta_1 = -i\eta_1\, ,
\ee
where we used the decomposition (\ref{4dim_decomposition_2}) of the 
ten-dimensional spinor representation. Comparing it with 
(\ref{varphi_theta}), we find that  the 
same fraction of supersymmetry can be preserved in the vacuum
if the $\theta$-angle vanishes, $\theta_{O3} = 0$. 

Let us repeat now the argument for $O9$-planes. An $O9$ 
is the fixed plane of the world-sheet parity, and therefore the 
invariant supersymmetry parameter $\epsilon_{I}$ should satisfy
\be
\epsilon_2 = \epsilon_1\, .
\label{09_susy1}
\ee
This condition is compatible with eq.~(\ref{susy_magnbrane}) for
\be
\epsilon_1 = \rho(\mathcal{F})\Gamma^{(11)}\epsilon_1\, .
\label{O9_susy2}
\ee
Using again eq.~(\ref{varphi_theta}) and choosing the chirality of 
the 
supersymmetry parameter in type IIB theory
to be positive, we get from (\ref{O9_susy2}) and (\ref{varphi_theta}) 
the 
condition on the $\theta$-angle, $\theta_{O9}= -\frac{\pi}{2}$. 

A similar analysis shows that the presence of (unmagnetized) 
$D5$ branes or $O5$ planes  is 
compatible with the choice for the 
$\theta$-angle, $\theta_{O9}= -\frac{\pi}{2}$, while the  
presence of $D7$ branes and $O7$ planes is compatible with the choice of the 
$\theta$-angle for $O3$ planes, $\theta_{O3} = 0$.


\section{Moduli stabilization}\label{sec:moduli_stab}

Following our analysis of eqs.~(\ref{complex_structure_condition1}) 
and (\ref{kaehler_condition2}), we have seen that a single magnetized 
$D9$ brane stack preserves $\mathcal{N}=1$ supersymmetry in four dimensions 
for a restricted closed string moduli space. As we see below, if we 
introduce several magnetic fluxes in the world-volume of different 
$D9$ branes, it will be possible to fix completely all moduli but 
the dilaton.


\subsection{Example: Orthogonal Torus}\label{subsec:orthogonal}

Before proceeding with the analysis of equations (\ref{kaehler_condition1}) 
and (\ref{complex_structure_condition1}), we consider for illustration a 
simple example of a compactification on a factorized four-dimensional 
orthogonal torus $T^4=(T^2)^2$. Thus, the non diagonal part of the 
K\"ahler form and complex structure $\tau$-matrix, as well as the real 
part of the diagonal elements $\tau_{ii}, \,\, i = 1,2$ are set equal to zero. 
The remaining moduli are given by the two $T^2$ volumes 
 \be
J_{1} =4\pi^2 R_{1}R_{2} \,\,\,\, , \,\,\, J_{2} =4\pi^2 R_{3}R_{4}
\ee
and the corresponding shapes
\be
\tau_{11} = i\frac{R_{2}}{R_{1}} \,\,\,\,, \,\,\, \tau_{22} = i 
\frac{R_{4}}{R_{3}},
\ee
where $R_{k}, \,\, \,\, k = 1, \dots 4 $ are the four radii:
$x_{i} = x_{i} + 2\pi R_{2i-1}$ and $y_{i} = y_{i} + 2\pi R_{2i}$, 
with $i=1, 2$. Following the Dirac quantization condition, 
the magnetic fields introduced in any 2-cycle of $T^{4}$ are quantized 
as follows:
\be
H_{kl}=2\pi\alpha'{F_{kl}\over V_{kl}} =  
\frac{m_{kl}}{n_{kl}}{\alpha'\over R_{k}R_{l}} =: 
p_{kl}\frac{\alpha'}{R_{k}R_{l}}\, \, ,\,\,\,\, p_{kl}\in \mathbb{Q}\, ,
\label{Hquant}
\ee
where $V_{kl}=4\pi^2R_kR_l$ is the area of the 2-cycle.

Consider now a $D9$ brane coupling to the  magnetic field $F$ 
which is switched on only in the two diagonal $T^2$-directions 
$F = (F_{x_{1}y_{1}},F_{x_{2}y_{2}})$. This magnetized brane 
preserves $\mathcal{N}=1$ supersymmetry in $d=6$ 
dimensions, when one of 
the scalars coming from the $d=10$ vector multiplet gets zero 
mass~\cite{Angelantonj:2000hi}. The corresponding fluxes then satisfy 
 $ |H_{x_{1}y_{1}}| = |H_{x_{2}y_{2}}|$. This condition can 
be understood as a condition on the K\"ahler form: 
\be
\frac{p_{x_{2}y_{2}}}{J_{2}} = \pm \frac{p_{x_{1}y_{1}}}{J_{1}}\, . 
\label{orthogonal_J}
\ee
On the other hand, if the field $F$ is switched on in the directions 
$F=(F_{x_{1}y_{2}},F_{x_{2}y_{1}})$, it preserves  $\mathcal{N}=1$ 
supersymmetry in $d=6$ when  $|H_{x_{1}y_{2}}| = |H_{x_{2}y_{1}}|$, 
which can now be understood as a condition on the complex structure 
moduli:
\be
p_{x_{1}y_{2}}\tau_{11} = \pm p_{x_{2}y_{1}}\tau_{22}\, .
 \label{orthogonal_t1}
\ee
Finally, if the $D9$ brane couples to a magnetic field $F = 
(F_{x_{1}x_{2}},F_{y_{1}y_{2}})$, the $\mathcal{N}=1$ supersymmetry in 
$d=6$ dimensions is recovered for $|H_{x_{1}x_{2}}| = 
|H_{y_{1}y_{2}}|$, which can also be written as a condition on the 
complex structure moduli: 
\be
p_{x_{1}x_{2}}\tau_{11}\tau_{22} = \pm p_{y_{1}y_{2}}\, .
\label{orthogonal_t2} 
\ee

The $D9$ brane action in the background given in (\ref{orthogonal_J}) reads
\be
S = - \int_{\mathcal{M}_{10}}T_{9}\sqrt{|g+H|} - 
\mu_{9}\int_{\mathcal{M}_{10}} C e^{H}\, .
\label{D9action}
\ee
$T_p$ and $\mu_p$ are the respective tension and R-R charge of a $Dp$ brane:
\be
T_{p}= {1\over g_s}{(\alpha')^{-{p+1\over 2}}\over (2\pi)^p}
\quad ;\quad \mu_p=g_s T_p\, ,
\ee
where $g_s=<e^\phi>$ is the string coupling, normalized so that 
for a $D3$ brane $g_s=2\alpha$, with $\alpha$ the usual 4d
gauge coupling.

After compactification to $d=6$ dimensions on the torus $T^{4}$, 
the action (\ref{D9action}) becomes for a supersymmetric configuration:
\bea
S &= &- \int_{\mathcal{M}_{10}}\left( T_9 \sqrt{g_{10}} +\mu_9 C_{10}\right)\\
  &  & -\int_{\mathcal{M}_6}( T_5\sqrt{g_6}\, 
|m_{1}m_{2}| + \mu_{5}m_{1}m_{2}C_{6}),
\label{action_orthogonal_6}
\eea
where we denote $m_i = m_{x_iy_i}$ and $n_i = n_{x_iy_i}$.
We therefore see that the magnetized $D9$ brane mimics 
the behavior of a $D5$ brane or a $\bar{D}5$ anti-brane, 
depending on the sign of $m_{1}m_{2}$~\cite{Angelantonj:2000hi}. 
We therefore have to introduce $O5$ or ${\bar O}5$ planes in order to 
cancel the induced tadpole. 
The cases where the fluxes satisfy (\ref{orthogonal_t1}) or 
(\ref{orthogonal_t2}) give similar results.

Let us consider now the compactification on a six-dimensional
factorized orthogonal torus
$T^{6}=(T^2)^3$ with moduli $J_i = 4\pi^2R_{2i-1}R_{2i}$ and 
$\tau_{ii} = i\frac{R_{2i}}{R_{2i-1}}$ for $i=1,2,3$.  The $\mathcal{N} = 1$ 
supersymmetry condition for a $D9$ brane coupled to a magnetic field 
$F = (F_{x_{1}y_{1}},F_{x_{2}y_{2}},F_{x_{3}y_{3}})$ can be 
deduced for example from the T-dual picture as a condition on the angles 
$\varphi_i = \arctan{H_{x_iy_i}}$:
\be
|\varphi_1|+|\varphi_2|-|\varphi_3|=0\, .
\label{ex}
\ee
This relation can be written in terms of the magnetic fluxes as 
\be
|H_{x_{1}y_{1}}| + |H_{x_{2}y_{2}}| - |H_{x_{3}y_{3}}| =  -|H_{x_{1}y_{1}}H_{x_{2}y_{2}}H_{x_{3}y_{3}}|\, , 
\label{orthogonal_susy_3}
\ee
when the wrapping numbers $n_i$ are positive, corresponding to the absence of antibranes, or equivalently to angles 
$|\varphi_i | \in [0,\pi/2]$. By the use of the Dirac quantization (\ref{Hquant}), 
this condition can also be understood as a condition on the K\"ahler moduli. 

In the above supersymmetric vacuum, the square root in the DBI 
action simplifies:
\be
V_{DBI} = - \int_{\mathcal{M}_{10}}T_9 \sqrt{g_{4}} \left(1- 
|H_{x_{1}y_{1}}H_{x_{2}y_{2}}|+
|H_{x_{2}y_{2}}H_{x_{3}y_{3}}|+
|H_{x_{1}y_{1}}H_{x_{3}y_{3}}|\right)\, .
\nonumber 
\ee
Thus, in this configuration, charges and tensions are induced in all three 2-cycles 
$[x_i,y_i],\, i=1,2,3$. However,  unlike the $T^4$ case  (\ref{action_orthogonal_6}),  
negative tension is induced in the 2-cycle $[x_3,y_3]$. 
This is important for making possible induced 5-brane tadpole cancellation in a
supersymmetric configuration involving only magnetized $D9$ branes,
{\em i.e.} in the absence of (anti-) $D5$ branes and
$O5$ planes, as it will become clear below.


\subsection{Conditions for $\mathcal{N}=1$ supersymmetry}\label{general}

Here, we perform a general study of eqs.~(\ref{kaehler_condition1}) and (\ref{complex_structure_condition1}), that can be interpreted as conditions
for fixing the moduli in terms of the magnetic fluxes.
For that, we consider $K$ stacks of $N_{a}$  $D9$ branes, with 
$a = 1, \cdots , K$. Furthermore, we introduce on each stack a 
background magnetic field with constant field strength $F^{a}$ on 
the corresponding
world-volume and charge $q_{a}$.  The magnetic fields are
separately quantized, following the Dirac condition
\be
q_a F^a_{kl} =2\pi\cdot \frac{m^{a}_{kl}}{n^{a}_{kl}}\equiv 2\pi\cdot 
p^{a}_{kl} \,\,\, , \  p^{a}_{kl} \in \mathbb{Q} \,\,\, , a = 1 , \cdots , K\, .
\label{diraca}
\ee

Written in the complex coordinates (\ref{complex_structure}), the  
field strength decomposes in a purely holomorphic and mixed part:
\bea
F^a_{(2,0)} & = & \!\!\!\! 2\pi {(\tau-\bar{\tau})^{-1}}^T \!\!\!\left[ \tau^{T} p^{a}_{xx} 
\tau - \tau^{T}{p^{a}_{xy}} - p^{a}_{yx}\tau + 
p^{a}_{yy}\right] (\tau-\bar{\tau})^{-1} \label{purelyholo}\\
F^a_{(1,1)} & = &\!\!\!\! 2\pi {(\tau-\bar{\tau})^{-1}}^T\!\!\! \left[ -\tau^{T} 
p^{a}_{xx}\bar{\tau} + \tau^{T}{p^{a}_{xy}} + p^{a}_{yx}\bar{\tau} - 
p^{a}_{yy}\right]\!\! (\tau-\bar{\tau})^{-1}
\eea
where the matrices $(p^{a}_{xx})_{ij}$, $(p^{a}_{xy})_{ij}$ and 
$(p^{a}_{yy})_{ij}$
enter in the quantized field strength (\ref{diraca}) in the directions
$(x^i,x^j)$, $(x^i,y^j)$ and $(y^i,y^j)$, respectively.
Note that $F^a_{(2,0)}$ and $F^a_{(1,1)}$ are $3\times 3$
matrices that correspond to the upper half of the matrix 
$\mathcal{F}^a$:
\be
\mathcal{F}^a=-2\pi i\alpha' \left(
\begin{array}{cc}
F^a_{(2,0)} & F^a_{(1,1)}\\
-{F^a}^\dagger_{(1,1)} & {{F^a}^*}_{(2,0)}\\
\end{array}
\right)\, ,
\label{matrixF}
\ee
which is the modified field strength in the 
cohomology basis $e_{i\bar{j}}$ defined in (\ref{e_basis}).
The supersymmetry conditions for each stack read:
\bea
F_{(2,0)}^{a} & = & 0 \,\,\, ,\  a = 1 , \cdots , K  
\label{complex_condition1a}\\
(iJ+\mathcal{F}^{a})^{3} & = & e^{i 
\theta_{a}}\sqrt{|g+\mathcal{F}^{a}|} \,\,\, ,\ a = 1 , \cdots , K\, . 
\label{kaehler_condition1a}
\eea
All  $\theta_{a}$'s have to be the same in order to preserve the same 
supersymmetry. We then have either $\theta_{a} = 
0,$ or $\theta_a = -\frac{\pi}{2}\,\,\, \forall\, a$.
Using eq.~(\ref{purelyholo}), the first condition 
(\ref{complex_condition1a}) can be seen as a restriction on the 
parameters of the complex structure matrix elements $\tau$:
\be
\tau^{T} p^{a}_{xx} \tau - \tau^{T}{p^{a}_{xy}} - p^{a}_{yx}\tau + 
p^{a}_{yy} = 0\, .
\label{M20_condition}
\ee
The second set of conditions (\ref{kaehler_condition1a})
gives, for fixed fluxes, restrictions on the K\"ahler parameters. 
Given the 
parametrization of the K\"ahler form (\ref{kaehler_structure}), these
conditions read:
\be
\tan{\theta}\left(J\wedge J \wedge \mathcal{F}^a - 
\mathcal{F}^a\wedge\mathcal{F}^a \wedge \mathcal{F}^a\right) = 
J \wedge J \wedge J - J\wedge 
\mathcal{F}^a \wedge \mathcal{F}^a\, .
\label{cond_kahlerX}
\ee
For vanishing purely holomorphic field strength imposed by 
(\ref{M20_condition}), the expression for 
 $\mathcal{F}$ reduces to the matrix 
\be
\mathcal{F}^a=\left(
\begin{array}{cc}
0 & Y^a\\
{Y^a}^\dagger & 0\\
\end{array}
\right)\quad ;\quad
Y^a=\frac{1}{2}(2\pi)^2 \alpha' {\rm{Im}} {\tau^{-1}}^T
\left( p_{yx}^a - \tau^{T} p^a_{xx}\right)\, .
\label{11part}
\ee
This splits in the real and imaginary parts:
\bea
{\rm{Re}} Y^a &=& \frac{(2\pi)^2\alpha'}{2}{\rm{Im}} 
{\tau^{-1}}^T\left( p^{a}_{yx} - 
{\rm{Re}}\tau^{T}p^{a}_{xx}\right)\, ,\\
{\rm{Im}}Y^a &=& -\frac{(2\pi)^2 \alpha'}{2}p^a_{xx}\, . 
\label{11part2}
\eea

Inspection of eqs.~(\ref{M20_condition}) and (\ref{cond_kahlerX}) 
shows that for each stack of magnetized $D9$ branes, we have up to 
three complex conditions 
for the moduli of the complex structure, depending on the directions 
in which the fluxes are switched on, whereas only one complex
condition can be set on the K\"ahler moduli. Therefore, to fix all 
K\"ahler 
moduli, we must add more stacks of branes compared to the ones 
needed to fix the same number of complex structure moduli. 


\subsection{Tadpole conditions}\label{sec:tadpoles}

The last information we need for a consistent setup is the number and 
location of lower dimensional objects, such as (anti-) $D$-branes or 
(anti-) orientifold planes, in order to cancel the 
tadpoles induced by the fluxes. In the case 
where the supersymmetry conditions 
(\ref{complex_structure_condition1}) and (\ref{kaehler_condition2}) 
are satisfied, the DBI and WZ actions read:
\bea
V_{DBI}  & = & -T_{9} \sum_{a=1}^{K}\,N_{a} \int_{\mathcal{M}_{10}^{a}}\sqrt{|g+\mathcal{F}^{a}|} 
\nonumber \\
         & = & -T_{9}\,\sum_{a=1}^{K}\,N_{a} 
\int_{\mathcal{M}_{4}}\sqrt{|g_4|}\int_{\mathcal{M}^{a}_{6}} 
 {\rm Re}\, [ e^{-i\theta_{a}}(iJ+\mathcal{F}^{a})^{3}] 
\nonumber \\
	  &=&  T_{9}\sum_{a=1}^{K}\,N_{a}\int_{\mathcal{M}_{4}}\sqrt{|g_4|}\int_{\mathcal{M}^{a}_{6}} 
         \bigg\{ \sin\theta_{a}\left(J\wedge J \wedge J - 
J\wedge\mathcal{F}^{a}\wedge\mathcal{F}^{a}\right)
\nonumber \\
& &+\,\, \cos\theta_{a}\left( J\wedge J \wedge \mathcal{F}^{a}-
\mathcal{F}^{a}\wedge \mathcal{F}^{a} \wedge \mathcal{F}^{a}\right)\bigg\}
\\
V_{WZ} & = & -\mu_{9}\sum_{a=1}^{K}\,N_{a}\int_{\mathcal{M}^a_{10}} C e^{\mathcal{F}^{a}} 
\nonumber \\
       & = & -\mu_{9}\sum_{a=1}^{K}\,N_{a}\int_{\mathcal{M}^a_{10}}
\bigg\{ C_{10}+C_{8}\wedge \mathcal{F}^{a} \\
       &  & +C_{6}\wedge \mathcal{F}^{a}\wedge \mathcal{F}^{a}
       + C_{4}\wedge \mathcal{F}^{a}\wedge \mathcal{F}^{a}\wedge \mathcal{F}^{a}
\bigg\} \, ,\nonumber
\eea
where the integral over the manifold $\mathcal{M}^{a}_6$ takes 
into account the winding numbers $n_{kl}$ of the different branes.

One can easily recognize in the above expressions the various 
tadpoles induced by the magnetic background fluxes. They correspond 
in the respective order to 10, 8, 6 and 4 dimensional spacetime filling 
tadpoles. In a consistent compactification, the introduction of $Dp$ branes 
(or $Op$ planes) with $p=3,5,7,9$ is then in general needed in order to cancel 
them term by term.  This is however not possible in a supersymmetric 
configuration, because the simultaneous presence of $Dp$ and 
$D(p+2)$ branes (as well as $Op$ and $O(p+2)$ planes) would break 
all supersymmetries. Thus, supersymmetry can be 
preserved only for the special values of the angles $\theta_{a}$ discussed 
in the previous section, namely 
$\theta_{a} = 0$ and $\theta_{a} = -\frac{\pi}{2}$. 

Indeed, when  $\theta_a = -\frac{\pi}{2}$, the magnetized $D9$ branes preserve the 
same supersymmetry as $O9$ planes and the DBI action reads
\bea
V_{DBI} &=&\!\!\! -T_{9}\sum_{a=1}^{K}N_{a} 
\int_{\mathcal{M}_{4}}\sqrt{|g_4|}\int_{\mathcal{M}^{a}_{6}}
(J\wedge J \wedge J
- J\wedge\mathcal{F}^{a}\wedge\mathcal{F}^{a})
\label{DBIO9}\\
V_{WZ} & = &\!\!\!  -\mu_{9}\sum_{a=1}^{K}N_{a}\int_{\mathcal{M}_{10}^{a}} \bigg\{ C_{10} + 
C_{6}\wedge\mathcal{F}^{a}\wedge\mathcal{F}^{a}) \bigg\}\, ,
\label{WZO9}
\eea
where in the last equation we assumed that the magnetized abelian generators
are embedded in higher dimensional non-abelian groups and are traceless.

It is now convenient to consider a real 
basis $\omega_r$ of $H^2(T^6)$, with $r = 1, \cdots , h_2$, in which  
the quantization condition (\ref{diraca}) for the magnetic fluxes reads:
\be
\frac{1}{2\pi} q_aF^a_r = \frac{m^a_r}{n^a_r} = p^a_r\, .
\ee
We also define the quantity
\be
\mathcal{K}_{rst} = \int_{T^6} \omega_r \wedge  \omega_s  \wedge\omega_t 
\label{Ksign}
\ee
which is a sign, following the orientation chosen in (\ref{orientation}). 
The 9-brane R-R charge, $q_{9, R}$, coming from the first term of 
(\ref{WZO9}), reads 
\be
q_{9, R}=\sum_{a=1}^{K}\, \sum_{r,s,t}\, N_a\, \mathcal{K}_{rst}n^a_r n^a_s 
n^a_t \, ,
\label{q9R}
\ee
while the corresponding contribution to the tension, $q_{9, NS}$, which 
comes from the first term of (\ref{DBIO9}), equals:
\be
q_{9, NS}=\sum_{a=1}^{K}\, \sum_{r,s,t}\, N_a\, |\mathcal{K}_{rst}n^a_r n^a_s 
n^a_t| \, .
\ee
Since we start with type I string theory with an $O9$ plane carrying 16 units 
of R-R charge and tension, the R-R tadpole cancellation condition implies 
$q_{9, R}=16$, while supersymmetry imposes that 
$\mathcal{K}_{rst}n^a_r n^a_s n^a_t $ is positive:
\be
q_{9, R}=16\quad ;\quad  \mathcal{K}_{rst}n^a_r n^a_s n^a_t \ge 0\,\, .\,\,
\forall a=1,\dots,K\, .
\label{tadpole_9}
\ee

The second set of  conditions comes from the induced 5-brane R-R charges 
and tensions, emerging from the second term of eqs.~(\ref{WZO9}) and 
(\ref{DBIO9}). For each 2-cycle $C^{(2)}_r$ of the torus $T^{6}$, 
they are given by $q_{5,R}^{r}$ and $q_{5,NS}^{r}$, respectively:
\bea
q_{5,R}^{r} & = &\sum_{a=1}^{K}\, \sum_{s,t}\, N_a\, \mathcal{K}_{rst}\, n^a_r m^a_s 
m^a_t =: \sum_{a=1}^{K}\, N_a\, q^{a}_{r}\, , 
\label{q5R}\\
q_{5,NS}^{r} & = -&\sum_{a=1}^{K}\, \sum_{s,t}\, N_a\, |\mathcal{K}_{rst}\, n^a_r| m^a_s 
m^a_t \quad ;\quad \forall\, r =1,\cdots, h_2\, .
\label{sum_tadpole5_NS}
\eea
To simplify our analysis, we will restrict ourselves to the ``standard" case with 
$\mathcal{K}_{rst}\, n^a_r\ge 0$ for any choice of 2-cycles $r$, $s$ and $t$.
This amounts to considering angles $|\varphi_i |\le \pi/2$, or
equivalently imposing the absence of anti-branes.
Then, the induced tensions $q_{5,NS}^{r}$ and charges 
$q_{5,R}^{r}$ have opposite signs, implying that we have to add 
$\bar{D}5$ branes and/or $\bar{O}5$ planes in order to cancel the induced tadpoles. 
But then, such a configuration would break all supersymmetries. 
It follows that the total 5-brane tadpole contribution must vanish for any 2-cycle $r$:
\be
q_{5,R}^{r}  =0 \,\,\, , \,\,\, \forall r= 1,\cdots, h_2\, .
\label{tadpole_5}
\ee

As a result, we will impose the R-R tadpole cancellation conditions 
(\ref{tadpole_9}) and (\ref{tadpole_5}): 
$q_{9, R}=16$ and $q_{5,R}^{r}  =0$, together with the supersymmetry
constraints (\ref{complex_condition1a}) or equivalently 
(\ref{M20_condition}), and 
\be
\mathcal{F}^a\wedge J \wedge J 
=\mathcal{F}^a\wedge\mathcal{F}^a\wedge \mathcal{F}^a, \,\,\, a = 
1,\ldots , K\, ,
\label{kaehler_condition_thetapi}
\ee
or equivalently $\sum_i\varphi_i^a=0$ (mod $2\pi$), for any magnetized $D9$ 
brane stack $a$, in the diagonal basis of section 4.2.

Furthermore, a condition of positivity for the real part of  
$\Phi_{a}$ defined in eq.~(\ref{Phi}) has to be satisfied for each $a$,
as it corresponds to the modified world-volume element of 
each separate brane stack:
\be
{\rm Re}(e^{-i\theta_a}\Phi_a) > 0\, , \quad \forall\, a 
= 1,\cdots , K\, ,
\ee
with
\be
\Phi_a = 
(iJ+\mathcal{F}^a)\wedge(iJ+\mathcal{F}^a)\wedge(iJ+\mathcal{F}^a)\, .
\label{Phia}
\ee
For $\theta_a=-\pi/2$, it reduces to the condition:
\be
J\wedge J\wedge J-J\wedge\mathcal{F}^a \wedge\mathcal{F}^a > 0\, ,
\label{Jcond}
\ee
which is equivalent to the positivity of the DBI lagrangian.

A similar analysis can be done for the choice of the angles
$\theta_{a}=0$, where the
magnetized $D9$ branes preserve the same supersymmetry as $O3$ planes.
This case can be obtained by T-duality from the above analysis, in a
straightforward way. The role of $n_r^a$ and $m_r^a$ are then interchanged
and the induced 5-brane R-R charge and tension is replaced by 7-brane ones.


\section{Explicit example with magnetized $D9$ branes
}\label{sec:example}

In this section, we present an explicit example of a toroidal model, 
where both complex structure and 
K\"ahler class moduli are completely fixed in terms of specific choices for 
magnetic fluxes.
Using the counting of conditions we presented in section 5.2, one has 
to introduce at least five brane stacks in order to fix all moduli. 
The model we describe below has instead nine stacks of branes with 
fluxes in different directions on their world-volume and $\theta^a= 
-\frac{\pi}{2}$. This corresponds to a type I string theory with 
constant internal magnetic fields on $D9$ branes. 

\subsection{Complex structure moduli stabilization}

The choice for the fluxes in the world-volume of the different 
magnetized $D9$ branes is given in Table 1.
Following the Dirac quantization (\ref{diraca}) and the condition 
(\ref{complex_condition1a}), the first three stacks of branes fix all 
complex structure moduli but $\tau_{12}$. The diagonal
elements are given by: 
\be
p^1_{x_1y_2}\tau_{11} = p^1_{x_2y_1}\tau_{22}\,\,\, ,\,\, \, p^2_{x_1y_3}\tau_{11} = p^2_{x_3y_1}\tau_{33}\,\,\, 
, \,\,\, p^3_{x_1x_2}\tau_{11}\tau_{22} = -p^3_{y_1y_2}
\label{6.1}
\ee
These conditions on the complex structure reproduce the results described 
in eqs (\ref{orthogonal_t1}) and  (\ref{orthogonal_t2}). They can be written as 
\be
\tau_{11}=i\sqrt{k_1\cdot k_3} \,\,\, , \,\,\, 
\tau_{22}=i\frac{\sqrt{k_3k_1}}{k_{1}}\,\, , \,\, 
\tau_{33}=i\frac{\sqrt{k_1 k_3}}{k_2},
\label{complex_fixed1}
\ee
where $k_i$ are the rational numbers
\be
k_1 = \frac{p^1_{x_2 y_1}}{p^1_{x_1 y_2}}\,\, , \,\, k_2 = 
\frac{p^2_{x_3 y_1}}{p^2_{x_1 y_3}} \,\, , \,\, k_3 = \frac{p^3_{y_1 
y_2}}{p^3_{x_1 x_2}}.
\label{ki}
\ee 
Moreover, all but $\tau_{12}$ off-diagonal elements vanish: 
$\tau_{ij}=0$.

\begin{table} \label{table1}
\vskip-0.5cm
\begin{center}
$$
\begin{array}{|c||c|c|c|}
\hline
\mathrm{Stack} \sharp & \mathrm{Fluxes} & \mathrm{Fixed\ moduli} & 
\mathrm{5-brane\ localization}\\
&&&\\
\hline
&&&\\
\sharp 1  & (F^1_{x_1 y_2},F^1_{x_2 y_1})& 
\tau_{31} = \tau_{32}=0 & [x_3,y_3]\\ 
          &                          &  p^1_{x_1 y_2}\tau_{11}= 
\tau_{22}p^1_{x_2 y_1} &                    \\
&&&\\
\hline
 &&&\\
\sharp 2  & (F^2_{x_1 y_3},F^2_{x_3 y_1})& 
\tau_{21} = \tau_{23}=0 &  [x_2,y_2]\\
          &                          & p^2_{x_1 y_3}\tau_{11}= 
\tau_{33}p^2_{x_3 y_1} & \\
&&&\\
\hline
 &&&\\
\sharp 3  & (F^3_{x_1 x_2},F^3_{y_1 y_2})& 
\tau_{13}=0& [x_3,y_3] \\
          &                          & 
\tau_{11}\tau_{22}=-\frac{p^3_{y_1 y_2}}{p^3_{x_1 x_2}} & \\
&&&\\
\hline 
\hline
&&&\\
\sharp 4  & (F^4_{x_2 x_3},F^4_{y_2 y_3})& 
\tau_{12}=0  & [x_1,y_1] \\
&&&\\
\hline
&&&\\
\sharp 5  & (F^5_{x_1 x_3},F^5_{y_1 y_3})&  & 
 [x_2,y_2]\\
&&&\\
\hline
&&&\\
\sharp 6  & (F^6_{x_2 y_3},F^6_{x_3 y_2})&  & 
 [x_1,y_1]\\
&&&\\
\hline
\end{array}
$$
\end{center}
\vskip-0.5cm
\caption{ Fixed complex structure moduli for each magnetized stack 
$\sharp$ of $ D9$ branes  depending on the quantized fluxes. The last 
column gives the localization on the 2-cycles $C^{(2)}_r$, with $r=[x_i,y_i]$, 
of the induced 5-brane charges, following section \ref{sec:tadpoles}.}
\end{table}
The last three stacks introduce new conditions on the complex moduli:
\be
\tau_{11}=ik_5\sqrt{\frac{k_6}{k_4}} \,\,\, , \,\,\, 
\tau_{22}=i\sqrt{k_4 
\cdot k_6}\,\, , \,\, 
\tau_{33}=i\frac{\sqrt{ k_4k_6}}{k_{6}},
\label{complex_fixed2}
\ee
where $k_i$ are the ratios:
\be
k_4 = \frac{p^4_{y_2 y_3}}{p^4_{x_2 x_3}}\,\, , \,\, k_5 = 
\frac{p^5_{y_1 y_3}}{p^5_{x_1 x_3}} \,\, , \,\, k_6 = \frac{p^6_{x_3 
y_2}}{p^6_{x_2 y_3}}\, .
\ee
Furthermore, they impose the vanishing of $\tau_{12}=0$.
Compatibility of the new conditions (\ref{complex_fixed2}) with those 
of eq.~(\ref{complex_fixed1}) imposes that the choice of fluxes is  
constrained to 
\be
|k_4| = \left|\frac{k_3}{k_2}\right| \,\,\, , \,\,\,  
|k_5| = \left|\frac{k_1k_3}{k_2}\right|\,\,\, , \,\,\, 
|k_6| =\left|\frac{k_2}{k_1}\right|\, .
\label{restriction_k}
\ee 


\subsection{K\"ahler class moduli stabilization}

From  the final form of the complex structure fixed to a diagonal 
purely imaginary matrix (\ref{complex_fixed1}), we can write the 
$\mathcal{F} _{(1,1)}$ part 
of the field strength as the matrix (\ref{11part}) with
\be
Y^a =-i\frac{(2\pi)^2 \alpha'}{2} (p_{xx}^a - \tau^{-1} 
p_{yx}^a )\, ,
\label{11part_fixed}
\ee
while the K\"ahler class is subject to the supersymmetry condition 
(\ref{kaehler_condition_thetapi}).
Let us denote for definiteness the volume of the 4-cycles, 
corresponding to $ J \wedge J$, as
\be
 (J \wedge J)_{i\bar{j}}=V_{i\bar{j}}\, ,
\ee
where the indices $i, \bar{j}$ correspond to the $(1,1)$-cycle 
perpendicular to the given 4-cycle. 

In the above notation, the fluxes introduced in the 
world-volume of the stacks $\sharp 1$, $\sharp 2$ and $\sharp 6$ 
contribute only to the real part of the complexified field strength 
$\mathcal{F}_{(1,1)}$. Using then the reality condition for the 
corresponding field strength  $\mathcal{F} _{i\bar{j}}= 
\mathcal{F}^\star_{j\bar{i}}$, one finds that the supersymmetry 
condition 
(\ref{kaehler_condition_thetapi}) on the K\"ahler moduli leads to:
\be
V_{1\bar{2}} - V_{2\bar{1}} = 0 \,\,\, , \,\,\, V_{1\bar{3}} - 
V_{3\bar{1}} = 0 \,\,\, , \,\,\, V_{2\bar{3}} - V_{3\bar{2}} = 0\, .
\label{fixed_kaehler1}
\ee
On the other hand, the magnetic fluxes on the  stacks $\sharp 3$, 
$\sharp 4$ and $\sharp 5$ 
contribute only to the imaginary part of $\mathcal{F}_{(1,1)}$, and 
the 
condition (\ref{kaehler_condition_thetapi}) on the K\"ahler moduli
imposes:
\be
V_{1\bar{2}} + V_{2\bar{1}} = 0 \,\,\, , \,\,\, V_{1\bar{3}} + 
V_{3\bar{1}} = 0 \,\,\, , \,\,\, V_{2\bar{3}} + V_{3\bar{2}} = 0\, .
\label{fixed_kaehler2}
\ee
Thus, all together, the conditions (\ref{fixed_kaehler1}) and 
(\ref{fixed_kaehler2}) give:
\be
V_{i\bar{j}}=0  \,\, ,\,\,  \mathrm{or\,\,\, equivalently}\,\, 
J_{i\bar{j}}=0  \,\,\,\, \forall\, i \neq j\, .
\ee
It follows that only three K\"ahler parameters remain unfixed,
corresponding to the diagonal part of the K\"ahler form. We must 
therefore introduce some new stacks of branes which add new 
conditions on the K\"ahler parameters, but preserve the values 
of the complex structure found previously. 

One possibility is displayed in Table 2. 
\begin{table}
\label{table2}
\begin{center}
$$
\begin{array}{|c||c|c|}
\hline
\mathrm{Stack} \sharp & \mathrm{Fluxes} & D5 \mathrm{branes \, \, 
localization }\\
\hline\hline
&&[x_1,y_1]\\
\sharp 7  & (F^7_{x_1 y_1},F^7_{x_2 y_2},F^7_{x_3 y_3})  & 
 [x_2,y_2]\\ 
          &                         &[x_3,y_3]\\
\hline
 &&[x_1,y_1]\\
\sharp 8  & (F^8_{x_1 y_1},F^7_{x_2 y_2},F^8_{x_3 y_3})  & 
  [x_2,y_2]\\
          &&[x_3,y_3] \\
\hline
 &&  [x_1,y_1]\\
\sharp 9  & (F^9_{x_1 y_1},F^9_{x_2 
y_2},F^9_{x_3 y_3})&  [x_2,y_2] \\
          &                          & [x_3,y_3]\\\hline 
\end{array}
$$
\end{center}
\caption{Additional stacks of magnetized $D9$ branes allowing the 
stabilization of the diagonal part of the K\"ahler form. The last 
column gives the localization on the 2-cycles $C^{(2)}_r$, 
with $r=[x_i,y_i]$, of the induced 5-brane charges, 
as explained in section \ref{sec:tadpoles}.}
\end{table}
The last three sets of fluxes introduced in the stacks $\sharp 7$, 
$\sharp 8$ and $\sharp 9$ give the following restrictions on the 
diagonal 
part of the K\"ahler form:
\be
\left(
\begin{array}{ccc}
\mathcal{F}^7_{1}&\mathcal{F}^7_{2}& 
\mathcal{F}^7_{3}\\
\mathcal{F}^8_{1}&\mathcal{F}^8_{2}& 
\mathcal{F}^8_{3}\\
\mathcal{F}^9_{1}&\mathcal{F}^9_{2}& 
\mathcal{F}^9_{3}
\end{array}
\right)
\left(
\begin{array}{c}
    J_{2}J_{3}\\
    J_{1}J_{3}\\
    J_{1}J_{2}
\end{array}
\right)
=\left(
\begin{array}{c}
    \mathcal{F}^7_{1}\mathcal{F}^7_{2}\mathcal{F}^7_{3}\\
    \mathcal{F}^8_{1}\mathcal{F}^8_{2}\mathcal{F}^8_{3}\\
    \mathcal{F}^9_{1}\mathcal{F}^9_{2}\mathcal{F}^9_{3}\\
\end{array}
\right)\, .
\label{linear_sys}
\ee
Here, we denote the K\"ahler moduli $J_{x_iy_i} = J_i$ and  the 
diagonal part of  fluxes, which are purely real, by 
$\mathcal{F}^a_{x_iy_i}=\mathcal{F}^a_{i}$. The solution of 
this linear equation is constrained to positive values for the 
moduli $J_{i}, \,\, i=1,2,3$, as they represent the volumes of 
the corresponding 2-cycles $[x_i,y_i]$. 


\subsection{Consistency conditions}

We have seen above that the introduction of nine magnetized stacks 
of  $D9$ 
branes allows us to stabilize the complex structure and K\"ahler 
class $T^6$-metric deformations
to three factorized orthogonal tori $T^{2}$:  
\be
\tau_{11}=i\sqrt{k_1\cdot k_3} \,\,\, , \,\,\, 
\tau_{22}=i\frac{\sqrt{k_3k_1}}{k_{1}}\,\, , \,\, 
\tau_{33}=i\frac{\sqrt{k_1 k_3}}{k_2}, \,\,\, \tau_{ij}=0
\label{complex_fixed3}
\ee
This solution is acceptable only if the following three additional 
restrictions are fulfilled:
\begin{itemize}

    
\item  In each two-torus $T^2$, the complex structure describes the 
shape of the torus,
\be
\tau = \frac{R_2}{R_1}e^{i\varphi},
\ee
where $R_{1,2}$ are the two radii and $\varphi$ is the opening 
angle of the $T^{2}$ relative to the horizontal axis, 
with $0 < \varphi <\pi $. A purely real complex 
structure would correspond to the degenerate case and makes therefore 
no sense. From eq.~(\ref{complex_fixed3}), we get:
\be
k_{i} > 0 \,\,\,, \,\,\, \forall i=1,2,3.
\label{restriction_complex}
\ee
Then, once the fluxes on the three first three stacks are 
chosen, the conditions (\ref{restriction_k}) restrict 
the fluxes that can be switched on in the next three stacks,
since $k_i,\, \, i=4,5,6$, are determined.
Moreover, a similar positivity argument based on 
(\ref{complex_fixed2}) implies that $k_i$ are also
positive for $i=4,5,6$:
\be
k_{i} > 0 \,\,\,, \,\,\, \forall i=4,5,6.
\label{restriction_complex2}
\ee

                                   
\item  The K\"ahler forms describing the volume of $T^{2}$'s
must be real and positive:
\be
J_i=J_{x_iy_i} > 0 \,\,\, , \,\, \forall i=1,2,3\, .
\label{restriction_kahler}
\ee

                                   
\item  The real part of the six-form $i\Phi^{a}$ defined in (\ref{Phia}) is a 
volume form for every stack of branes and must therefore be positive, 
as displayed in eq. (\ref{Jcond}):
$J \wedge J \wedge J - J \wedge \mathcal{F}^{a} \wedge 
\mathcal{F}^{a}  > 0$ for all $a=1,\dots , 9$.
Consider the example of the first stack. Here, the only non vanishing
component of the expression (\ref{Jcond}) is written as:
\be
J_{x_{1}y_{1}}J_{x_{2}y_{2}}J_{x_{3}y_{3}} + 
J_{x_{3}y_{3}}\mathcal{F}_{x_{1}y_{2}}\mathcal{F}_{x_{2}y_{1}} > 0 \, .
\label{Jcond2} 
\ee
From the restriction (\ref{restriction_complex}), we find that the product of fluxes 
$p_{x_{1}y_{2}}p_{x_{2}y_{1}}$ is positive. From the restriction 
(\ref{restriction_kahler}), the K\"ahler forms $J_{x_{i}y_{i}}$ are also 
positive and the condition (\ref{Jcond2}) is therefore trivially satisfied. 
This argument can be similarly applied for all the first 
six stacks, but the situation is different for the last three stacks
of branes, where we have switched on fluxes in the three 
2-cycles $[x_{i},y_{i}],\,\, i=1,2,3$. The condition (\ref{Jcond}) can now 
be written as:
\be
1 - H^{b}_{1}H^{b}_{2} - H^{b}_{1}H^{b}_{3} - H^{b}_{2}H^{b}_{3} > 0 
\,\,\,, \,\,\, b=7,8,9 \, , 
\nonumber
\ee
where we defined the magnetic fields 
$H^{b}_{i} = \frac{\mathcal{F}^{b}_{x_{i}y_{i}}}{J_{x_{i}y_{i}}}$. 
Using the supersymmetry condition (\ref{kaehler_condition_thetapi}),
\be
H^{b}_{1} + H^{b}_{2} + H^{b}_{3} = H^{b}_{1}H^{b}_{2}H^{b}_{3} \, 
\,\,\, , \,\,\, b=7,8,9 \, ,
\ee
we get for instance (for $H_3^b\ne 0$)
\be
-\frac{H^{b}_{1}+H^{b}_{2}}{H^{b}_{3}}(1+(H^{b}_{3})^{2}) >0\,\, , 
\ee
which leads to the conditions on the fluxes:
\be
\bigg\{
\begin{array}{l}
    H^{b}_{1}+H^{b}_{2} > 0 \\
    H^{b}_{3} < 0
\end{array}
\,\,\,
\rm{or} \,\,\,
\bigg\{
\begin{array}{l}
    H^{b}_{1}+H^{b}_{2} < 0 \\
    H^{b}_{3} > 0
\end{array}\, .
\label{positivity_final}
\ee

\end{itemize}

\subsection{Tadpoles}

Above, we found a set of consistency conditions which restrict the 
allowed choice of background fluxes. They came from the geometrical 
character of the complex structure and K\"ahler class moduli, as well 
as from the positivity of the modified volume forms ${\rm{Re}}(i\Phi_{a})$. 
A second set of restrictions comes from the tadpole cancellation,
which we examine here. 

The nine stacks of magnetized $D9$ branes induce 5-brane 
R-R charges $q_{R}^{r}$ as well as NS-NS 
tensions in the three 2-cycles $r=[x_{i},y_{i}]$, 
$i=1,2,3$, given by the eqs.~(\ref{q5R}) and (\ref{sum_tadpole5_NS}).
As we argued in section 5.3, the absence of anti-branes implies
that the winding numbers are positive and the total R-R 5-brane
charge must vanish:
\be
\mathcal{K}_{rst}n^{a}_r\ge 0\qquad ;\qquad
\sum_{a=1}^{9}N_{a}q_{r}^{a} = 0 \,\,\, ,\,\,\, \forall 
r=[x_{i},y_{i}]\, ,
\label{tadpole_5_1}
\ee
where
\be
q_{r}^{a} = \sum_{st}\mathcal{K}_{rst}n^{a}_{r}m^{a}_{s}
m^{a}_{t} = \sum_{st}\mathcal{K}_{rst}n^{a}_{r}n^{a}_{s}
n^{a}_{t}p^{a}_{s}p^{a}_{t}\,\,\, , \,\,\, a=1,\dots,K= 9\, .
\label{tadpole_fluxes}
\ee
Moreover, the tadpole cancellation for the 9-brane R-R charge
$q_{9,R}$ of eq.~(\ref{q9R}) and tension leads to $q_{9,R}=16$
and the winding number positivity condition 
$\mathcal{K}_{rst}n^{a}_r n^{a}_s n^{a}_t\ge 0$, as displayed
in eq.~(\ref{tadpole_9}).

Let us analyze first the contributions of the first six stacks to 
(\ref{tadpole_5_1}). Following our choice for the magnetic fluxes,
the tadpoles are induced only in one of the three 2-cycles 
$[x_{i},y_{i}]$, as presented in Table 1. 
On the other hand, the positivity conditions
(\ref{restriction_complex}) and (\ref{restriction_complex2}) on the first six stacks of magnetized $D9$ branes 
restrict the sign of the product of fluxes $p^{b}_{s}p^{b}_{t} >0$, 
for $b=1,\dots,6$. All charges $q^b_{r}$ are therefore 
given by $q^b_{r} = 
\mathcal{K}_{rst}n^{b}_{r}n^{b}_{s}n^{b}_{t}p^{b}_{s}p_{b}^{t}$,
and are positive, following (\ref{tadpole_9}). 

The contributions of the last three stacks of branes must therefore cancel 
the positive charges $q^b_{r}$ induced by the first six stacks. 
In this case, the fluxes are 
switched on in the three 2-cycles $[x_i,y_i]$, $i=1,2,3$, and therefore 
tadpoles are induced in all these 2-cycles. Furthermore, the 
positivity condition (\ref{positivity_final}) reduces the possible 
signs of the induced R-R 5-brane charges $q_{[x_i,y_i]}^b$.
As the components $J_{x_{i}y_{i}}$ of the K\"ahler form have 
to be positive from the consistency requirement
(\ref{restriction_kahler}), a necessary condition for 
(\ref{positivity_final}) to be satisfied is to use different signs
for the fluxes. There are indeed three possibilities:
\be
\begin{array}{cc}
{\rm{sign}}(p^c_{x_1y_1}) = {\rm{sign}}(p^c_{x_2y_2}) \ne 
{\rm{sign}}(p^c_{x_3y_3})&  \\
{\rm{sign}}(p^c_{x_1y_1})\ne {\rm{sign}}(p^c_{x_2y_2}) = 
{\rm{sign}}(p^c_{x_3y_3}) &  \\
{\rm{sign}}(p^c_{x_1y_1}) = {\rm{sign}}(p^c_{x_3y_3}) \ne 
{\rm{sign}}(p^c_{x_2y_2}) & 
\end{array};\,\,\,\, c=7,8,9\, .
\label{flux_sign_7-9}
\ee
Using (\ref{tadpole_fluxes}) and the positivity condition 
(\ref{tadpole_9}), the  signs of the 5-brane charges 
$q_{[x_i,y_i]}^c$ are given by:
\be
{\rm{sign}}(q_{[x_i,y_i]}^c) = 
{\rm{sign}}(p^c_{x_jy_j}p^c_{x_ky_k}), \,\,\, i\ne j \ne k\ne i 
\,\,\, , \,\,\, c=7,8,9\, .
\label{flux_sign_q7-9}
\ee
We therefore notice that if we switch on fluxes in all 2-cycles 
$[x_i,y_i]$, two of the induced charges $q_{[x_i,y_i]}^c$ contribute 
negatively to the tadpoles whereas only one contributes positively. 
On the other hand, if one of the fluxes  $p_{x_iy_i}^c$ vanishes, 
the only tadpole contribution is precisely $ q_{[x_i,y_i]}^c$, which 
has to be negative, because of the supersymmetry condition 
(\ref{kaehler_condition_thetapi}). Thus, in both cases, we see that the 
5-brane charges induced by the last three stacks can in principe cancel the 
positive contribution of the first six stacks. 
Finally, the 9-brane charge $q_{9,R}$ (\ref{q9R}) receives only 
positive contributions from the brane wrapping numbers $n^a_r$
and the tadpole cancellation condition (\ref{tadpole_9})
can be easily satisfied. In Appendix A, we 
give an explicit numerical example where this is indeed realized. 


\section{Large dimensions}\label{sec:led}

We have seen above that there exist configurations of magnetized 
$D9$ branes which lead to $\mathcal{N}=1$ supersymmetric vacua 
in $d=4$ dimensions with all moduli fixed to the geometry of three 
factorized orthogonal torii. The non-vanishing complex structure 
matrix elements correspond then to the ratio of the radii within each 
of the three 2-cycles, $\tau_{ii} = i\frac{R_{2i}}{R_{2i-1}}$, whereas 
the K\"ahler class moduli describe their volumes, 
$J_i =4\pi^2 R_{2i-1}R_{2i}$. The latter are  functions of the fluxes 
switched on in the last three stacks of branes. On the contrary, the 
values of the complex structure moduli are determined only by the 
fluxes introduced in the world-volume of the first six stacks. 
Here, we analyze the possibility of inducing a hierarchy between 
the string length $l_s = \sqrt{\alpha'}$ and the values of the 
radii~\cite{ld}.

To this end, let us consider the generic configuration (\ref{linear_sys}) 
written in the real basis $x_i,y_i$, $i=1,2,3$, with given fluxes 
$p^c_{x_iy_i}$ such that $\mathcal{F}^c_{i} = 4\pi^2 \alpha' p^c_{x_iy_i}$, $c=7,8,9$. 
Moreover, we start from some given solution of the system that leads to
a consistent stabilization of the K\"ahler moduli, of order say 
$\alpha'$. This means that there are some positive $(J_{1},J_{2},J_{3})$ 
which solve the equations (\ref{linear_sys}), while
the induced 5-brane tadpoles in the three 2-cycles $[x_i,y_i]$:
\be
\left(
\begin{array}{c}
   q_{[x_1,y_1]} \\
   q_{[x_2,y_2]} \\
   q_{[x_3,y_3]} \\
\end{array}
\right)=
 \left(
\begin{array}{ccc}
m^7_{2}m^7_{3}&m^8_{2}m^8_{3}&m^9_{2}m^9_{3} \\
m^7_{1}m^7_{3}&m^8_{1}m^8_{3}&m_{1}m^9_{3}\\
m^7_{1}m^7_{2}&m^8_{1}m^8_{2}&m^9_{1}m^9_{2}
\end{array}
\right)
\left(
\begin{array}{c}
    N_7\\
    N_8\\
    N_9
\end{array}
\right)
\label{tadpole_sys}
\ee
are negative, and can be cancelled by the charges induced by the 
first six stacks. Here, we defined $q_r\equiv \sum_{a=7}^9 N_a q^a_r$,
$m^a_i\equiv m^a_{x_i y_i}$, and we have chosen for simplicity all 
winding numbers equal to one, $n^a_{x_i y_i}=1$ for $a=7,8,9$, 
so that the flux quanta $p^c_r$ become equal to $m^c_r$. 

Let us now rescale the three fluxes in the direction $[x_1,y_1]$ by a 
factor $\Lambda$. The conditions (\ref{linear_sys}) on the K\"ahler 
moduli become 
\be
\left(
\begin{array}{ccc}
\Lambda\mathcal{F}^7_{1}&\mathcal{F}^7_{2}& 
\mathcal{F}^7_{3}\\
\Lambda\mathcal{F}^8_{1}&\mathcal{F}^8_{2}& 
\mathcal{F}^8_{3}\\
\Lambda\mathcal{F}^9_{1}&\mathcal{F}^9_{2}& 
\mathcal{F}^9_{3}
\end{array}
\right)
\left(
\begin{array}{c}
    J_{2}^\Lambda J_{3}^\Lambda\\
    J_{1}^\Lambda J_{3}^\Lambda\\
    J_{1}^\Lambda J_{2}^\Lambda
\end{array}
\right)
=\left(
\begin{array}{c}
   \Lambda \mathcal{F}^7_{1}\mathcal{F}^7_{2}\mathcal{F}^7_{3}\\
   \Lambda \mathcal{F}^8_{1}\mathcal{F}^8_{2}\mathcal{F}^8_{3}\\
   \Lambda \mathcal{F}^9_{1}\mathcal{F}^9_{2}\mathcal{F}^9_{3}\\
\end{array}
\right),
\label{linear_sys_rescaled}
\ee
which have obviously as solution the rescaled K\"ahler form $J^\Lambda$:
\be
J^\Lambda=(\Lambda J_{1},J_{2},J_{3})\, .
\ee
Thus, we have triggered a hierarchy between the first component 
$J_{1}$ of the K\"ahler form and $\alpha'$ by a factor $\Lambda$,
given by the corresponding rescaling of the fluxes 
$m_{1}^c\to m_{1}^{c,\Lambda}= \Lambda m_{1}^c$ 
switched on in the 2-cycle $[x_1,y_1]$ of the first torus, for the three 
stacks of branes $c=7,8,9$. In this example, the other components 
of the K\"ahler form remain at their original values, of order $\alpha'$. 

The rescaling of the flux quanta induces also a rescaling 
$q^\Lambda_r$ of the 5-brane charges (\ref{tadpole_sys}):
\be
\left(
\begin{array}{c}
   q_{[x_1,y_1]}^\Lambda \\
   q_{[x_2,y_2]}^\Lambda \\
   q_{[x_3,y_3]}^\Lambda \\
\end{array}
\right)
=
\left(
\begin{array}{c}
   q_{[x_1,y_1]} \\
  \Lambda q_{[x_2,y_2]} \\
  \Lambda q_{[x_3,y_3]} \\
\end{array}
\right).
\label{tadpole_sys2}
\ee
This contribution must be cancelled by an appropriate rescaling of 
the fluxes introduced in the first six stacks. For simplicity, we 
consider the case where the ratio of the radii within each $T^2$
$R_{2i}/R_{2i-1} = -i\tau_{ii}$ remains of order one. The ratio of 
fluxes in each stack must therefore be of order unity, $k_a \sim 1$ for
$a=1,\dots,6$. Choosing all non-trivial winding numbers 
equal to one, ${\mathcal K}_{rst}n^a_r=1$, we get 
that the corresponding induced 5-brane charges (\ref{tadpole_fluxes}) 
are given by:
\be
\sum_{a=1}^6 \left(
\begin{array}{c}
q_{[x_1,y_1]}^{a,\Lambda} \\
q_{[x_2,y_2]}^{a,\Lambda} \\
q_{[x_3,y_3]}^{a,\Lambda} \\
\end{array}
\right)
=\left(
\begin{array}{c}
N_4 (m^{4,\Lambda}_{x_2x_3})^2 +N_6 (m^{6,\Lambda}_{x_2y_3})^2\\
N_2 (m^{2,\Lambda}_{x_1y_3})^2 +N_5 (m^{5,\Lambda}_{x_1x_3})^2\\
N_1 (m^{1,\Lambda}_{x_1y_2})^2 +N_3 (m^{3,\Lambda}_{x_1x_2})^2\\
\end{array}
\right)
\stackrel{!}{=}
-\left(
\begin{array}{c}
q_{[x_1,y_1]} \\
\Lambda q_{[x_2,y_2]} \\
\Lambda q_{[x_3,y_3]} \\
\end{array}
\right)\, ,
\label{tadpole_cancellation}
\ee
where $m^{a,\Lambda}_r$ denote the rescaled magnetic fluxes,
that we determine now. 

Indeed, the second part of 
eq.~(\ref{tadpole_cancellation}) follows from the 5-brane tadpole 
cancellation condition (\ref{tadpole_5_1}). It can be satisfied, 
if the fluxes of the stacks $\sharp 1$, $\sharp 2$, $\sharp 3$ and 
$\sharp 5$ are rescaled as:
\be
\begin{array}{cc}
m^{1,\Lambda}_{x_1y_2} =\sqrt{\Lambda} m^1_{x_1y_2}\,\,\,& , 
\,\,\,\,\,m^{2,\Lambda}_{x_1y_3} =\sqrt{\Lambda}  m^2_{x_1y_3}\, , \\
&\\ m^{3,\Lambda}_{x_1x_2} = \sqrt{\Lambda} m^3_{x_1x_2} \,\,\,& , 
\,\,\,\,\, m^{5,\Lambda}_{x_1x_3} = \sqrt{\Lambda} m^5_{x_1x_3}\, ,
\end{array}
\ee
while keeping the corresponding ratios $k_a$, for $a=1,\dots,6$,
fixed (of order unity). Moreover, the fluxes in the stacks $\sharp 4$ 
and $\sharp 6$ remain unchanged. An explicit numerical example 
is presented in Appendix \ref{app:led}.
It follows that there is an infinite but discrete class of supersymmetric 
solutions, where some radii can be made arbitrarily large.


\section{Stabilization of the R-R moduli}\label{sec:RR}

We have shown above that it is possible to fix all K\"ahler and 
complex structure moduli in a consistent toroidal 
compactification with magnetic fluxes, preserving 
$\mathcal{N}=1$ supersymmetry in four dimensions. We 
want now to address the question of stabilization of the R-R
moduli coming from the R-R 2-form. Recall that in the type I 
framework we work, the only R-R potential of type 
IIB string theory which survives the orientifold projection is the 
2-form $C_{2}$. Upon toroidal compactification from ten to four 
dimensions, it gives rise to  $\frac{6\times 5}{2}=15$ 
 massless scalars $c_{mn}$, with $m,n = 4,\ldots , 9$. Nine of them  
complexify the K\"ahler form $J$, whereas the remaining six form 3 
additional complex scalars. 
 
In a general magnetic background, we have shown that the 
K\"ahler moduli get fixed. 
Since the new vacuum preserves $\mathcal{N}=1$ supersymmetry, the 
real part of the K\"ahler moduli formed by the corresponding nine 
components of the R-R 2-form should be fixed, as well. To describe 
the stabilization mechanism, we remind first some properties of 
$C_{2}$. In the low energy effective action of the $\mathcal{N}=1$, 
$d=10$ type I string theory, the field strength $H_{3}$ of the $C_{2}$ 
potential, $H_3=dC_2$, gets 
modified by a Chern-Simons (CS) term coming from the gauge sector:
\be
S_{cl} = \frac{1}{2\kappa_{10}^{2}} \int e^{-2\phi}(- R\star1 +4d\phi 
\wedge \star d\phi -\frac{e^{2\phi}}{2}\tilde{H}_{3}\wedge \star 
\tilde{H}_{3})\, ,
\label{typeIeffective_action}
\ee
where $\kappa_{10}^{2} = \frac{1}{4}(2\pi)^{7}{\alpha'}^{4} $ and
\be
\tilde{H}_{3} = dC_2 - 
\frac{\alpha'}{4}{\rm Tr}(A\wedge dA -\frac{2i}{3} A 
\wedge A \wedge A)\, 
\label{modified_h3}
\ee
with $A$ the gauge potential. 
Note that $C_{2}$ appears in the 
action (\ref{typeIeffective_action})
only through derivatives, as expected. In fact, upon compactification 
in four dimensions, all R-R scalars are  associated to perturbative 
Peccei-Quinn symmetries. 

Consider now some $U(1)$ directions with constant internal magnetic 
field backgrounds. The corresponding gauge potential is
$A= (A_{\mu},A_m)$, where  $A_\mu$ is the four-dimensional field, 
with $\mu = 0,\ldots, 3$, and  $A_m$ describes the magnetic 
background potential given in section \ref{subsec:gauge_field}. 
In this case, the CS action simplifies and the interaction terms of the 
action (\ref{typeIeffective_action}) involving $C_2$ give rise to
\be
S_{int} = \frac{1}{4g_{10}^{2}}\sum_a^K 
N_a\int_{\mathcal{M}_{4}} dC_{2} \wedge \star (A^a\wedge F^a)\, ,
\ee
where $g_{10}^{2}=4\kappa_{10}^{2}/\alpha'$,
$F^a$ is the background field strength and $N_a$ is the 
number of branes in the $a$-th stack. We thus obtain a kinetic 
mixing term between the axion fields $c_{mn}$ and the $U(1)$ 
gauge bosons~\cite{Angelantonj:2000hi}:
\be
 S_{int} = \frac{1}{4g_{10}^{2}}\sum_a^K N_a 
\int_{\mathcal{M}_{4}} d^{4}x \sqrt{g_4}\,
 A^a_{\mu}\partial^{\mu}c_{mn}\,g_{{mn,kl}}\,F^a_{kl}\, ,
\label{A_mass}
\ee
where $g_{mn,kl}$ is the metric on the cohomology 
$H^{2}(T^{6})$, 
\be
g(\alpha, \beta)_{H^2} = \int_{T^{6}} \alpha \wedge \star \beta \,\,, 
\quad 
\forall\, \alpha, \beta \in  H^{2}(T^{6})\, .
\label{H2metric}
\ee

In the configuration presented in  section \ref{sec:example}, we 
introduced nine stacks of  magnetized $D9$ branes, with constant 
field strengths on their world-volume given in Tables 1 and 2. From 
the kinetic mixing  (\ref{A_mass}), it is easy to see that precisely 
nine combinations of the axion fields $c_{mn}$ are absorbed by the 
nine magnetized $U(1)$ gauge fields and form massive gauge 
multiplets, together with the corresponding K\"ahler moduli.  

In fact, in the complex basis (\ref{complex_structure}), they are
the nine R-R moduli with mixed indices 
$c_{i\bar{j}}$ that get absorbed into the mass of the gauge fields. 
More precisely, the 
cohomology $H^2$ splits into the cohomology $H^{2,0}\otimes 
H^{1,1}\otimes H^{0,2}$ and the metric (\ref{H2metric}) splits also 
in three terms corresponding to a metric $g_1$ on 
$H^{1,1}$ ~\cite{Strominger:1985ks},
\be
g_{1}(\alpha_{1},\alpha_{2}) = 
\frac{\int_{T^{6}}\alpha_{1}\wedge J \wedge J 
\int_{T^{6}}\alpha_{2}\wedge J \wedge J}{\int_{T^{6}}J \wedge J \wedge 
J } - \int_{T^{6}}\alpha_{1}\wedge \alpha_{2} \wedge J\, ,
\ee
a metric $g_2$ on $H^{2,0}$ and a metric $g_3$ acting on their product as:
\be
g_3 (\alpha, \beta) =  \int_{T^{6}} \alpha \wedge \star \beta 
\,\,, \quad 
\forall\, \alpha \in  H^{1,1}(T^{6}) \,\,\, \textrm{and} \,\,\, 
\beta  \in  H^{2,0}(T^{6})\, .
\ee 
Since the internal manifold has been fixed by the fluxes to be a product 
of three orthogonal $T^2$, the third part of the metric decomposition 
vanishes. Indeed, since the non vanishing elements of the metric are
of mixed type, $g_{i{\bar j}}$, the Hodge star operation maps 
$(2,0)$-forms into $(1,3)$-forms. It follows that the wedge product
$\alpha\wedge\star\beta$ is zero.
Moreover, the vacuum configuration requires a vanishing 
purely holomorphic field strength $F^a_{2,0}=0$
and, thus, the metric $g_2$ is irrelevant. Therefore, by the
form of the mixing (\ref{A_mass}), only the 
R-R moduli that are elements of the cohomology $H^{1,1}$ are absorbed 
into the mass terms of the $U(1)$ gauge fields. 

The purely holomorphic (and anti-holomorphic) part of the R-R 2-form 
can not be absorbed into the $U(1)$ mass terms. They enter however in 
the complex structure moduli. Indeed, the latter are 
parametrized by the matrix $\tau$ which has 
nine complex entries. Only six of them correspond to the purely 
(anti-) holomorphic variation of the metric ($\delta 
g_{\bar{i}\bar{j}}$) $\delta g_{ij}$. The remaining  three complex 
elements  correspond to the (anti-) holomorphic variation of the RR 
2-form ($\delta c_{\bar{i}\bar{j}}$) $\delta c_{ij}$, which is a
peculiarity of the toroidal compactification~\cite{Kachru:2002he}. 
These are therefore fixed by the conditions (\ref{complex_condition1a})
or equivalently (\ref{M20_condition}). As a result, all R-R moduli 
are also stabilized by our choice of magnetic fluxes.

\section{Generalizations and concluding remarks}\label{sec:CY}

In this section, we discuss the generalization of the method 
presented above to stabilize moduli in type I string compactifications 
preserving ${\mathcal N}=1$ supersymmetry in four dimensions,
such as orbifolds and Calabi-Yau manifolds. An immediate problem that 
arises in compactifications with strict $SU(3)$ holonomy is that the only 
cohomology which exists is that of 3-forms $H^{3,0}$ and $H^{2,1}$, 
as well as the one of 2-forms $H^{1,1}$. 
In other words, the cohomology $H^{2,0}$ is trivial. 
One could then naively conclude that the condition 
(\ref{complex_condition1a}) is trivially satisfied and, thus, 
the complex structure moduli cannot be fixed.

One possibility to overcome this difficulty is to combine the above 
mechanism with the
presence of NS-NS and R-R 3-form fluxes, that fix precisely the 
complex structure moduli and
the dilaton, but not the K\"ahler class. The latter can be 
fixed by turning on
internal magnetic fields on the $D9$ branes, leading to the condition 
(\ref{kaehler_condition1a}), or equivalently to (\ref{cond_kahlerX}).
The two mechanisms are therefore complementary.
Of course, in ${\mathcal N}=1$ compactifications with holonomy 
different than
$SU(3)$, such as $SU(2)$ times a discrete group factor, the 
cohomology $H^{2,0}$ is in general 
non-trivial and magnetic fluxes on holomorphic 2-cycles can be 
used to fix (at least part of) the complex structure moduli.
This is for instance the case of Enrique surfaces in ${\mathcal N}=2$ 
compactifications with vanishing Euler number~\cite{Ferrara:1995yx}.

On the other hand, below we point out that the above
conclusion is naive and our method can be used to fix the 
complex structure, even in compactifications with $SU(3)$ 
holonomy and trivial $H^{2,0}$ cohomology. The reason is that 
despite the absence of four-dimensional zero modes for the internal 
components of gauge fields, that transform non-trivially
under $SU(3)$ holonomy, one is still allowed to turn on (discrete)
internal magnetic fields that are invariant under the holonomy group.

To illustrate this point, we consider for simplicity a
${\mathcal N}=2$ supersymmetric example of the orbifold 
compactification $T^2 \times T^4/\mathbb{Z}_2$~\cite{{Gimon:1996rq}}. 
Let us denote the complex coordinate on $T^2$ by $z_1$,  and those on 
$T^4$ by $z_a $, with $a=2,3$. The $\mathbb{Z}_2$-orbifold acts on 
the coordinates as a parity operation on $T^4$:
\be
\mathbb{Z}_2: z_2 \rightarrow -z_2\,\,\,\,\, , \,\,\,\,\,z_3 
\rightarrow -z_3\, .
\ee
The dimension of the untwisted moduli space is therefore reduced. 
Only five untwisted complex structure moduli remain, which are 
denoted, following the notation (\ref{complex_structure}), by:
\be
 \tau_{11}\,\,\,\, , \,\,\, \tau_{ab} \quad ;\quad a,b = 2,3\, .
\ee
Similarly, the dimension of the  untwisted K\"ahler class moduli is 
reduced to five. 
Moreover, the purely holomorphic 2-forms of the type $dz^1\wedge 
dz^a$ are projected out by the orbifold action. The only remaining 
holomorphic 2-form is  $dz^2\wedge dz^3$. 
The massless open string spectrum on the $D9$ branes consists of a 
$\mathcal{N}=2$ vector multiplet of the gauge group $U(16)$ and of 
hypermultiplets transforming under the antisymmetric representation 
$\mathbf{120} + \overline{\mathbf{120}}$. The (complex) scalar 
components of the vector multiplets correspond to $A^e_{1}$, where 
$A$ is the gauge potential and the superscript $e$ denotes the 
$U(16)$ $\mathbb{Z}_2$-invariant (even) generators of the 
ten-dimensional $SO(32)$ gauge group.
On the contrary, the $\mathbb{Z}_2$ action on the group index $o$ 
of the two complex scalars $A^o_{a}$ of the hypermultiplets, is odd. 

For a purely holomorphic field strength to survive the orbifold 
projection, its components 
\be
F_{12}= \partial_{[1}A_{2]}\,\,\, ,\,\, F_{13}= 
\partial_{[1}A_{3]}\,\,\, ,\,\, F_{23}= \partial_{[2}A_{3 ]},
\ee
must have an even $\mathbb{Z}_2$-parity. It follows that $F_{12}$ and 
$F_{13}$
should involve internal derivatives of the gauge potential $A^o$ 
along the broken generators of the $SO(32)$, transforming as the 
hypermultiplets in the antisymmetric representation of $U(16)$:
\be
F_{12}= \partial_{[1}A^o_{2]}\,\,\, ,\,\, F_{13}= 
\partial_{[1}A^o_{3]}\, .
\ee
Thus, even if there is only one (2,0)-form $dz^2\wedge dz^3$ 
invariant under the 
$\mathbb{Z}_2$-orbifold parity, all three internal components of 
$F_{2,0}$ survive the orbifold action. 
In fact, following our analysis of section~\ref{sec:example}, they 
are all needed in order to fix the complex structure moduli.

In conclusion, in this work we have shown that the K\"ahler class and 
complex structure moduli
of type I string compactifications can be fixed by a suitable choice 
of internal magnetic fields, in
vacua preserving ${\mathcal N}=1$ supersymmetry in four dimensions.
Moreover, the moduli can be fixed at arbitrarily large values
by tuning the rational quantized fluxes.
In contrast to other stabilization mechanisms based on 3-form fluxes 
from the closed string sector, this method has an exact string 
description and, thus, a validity beyond the low-energy supergravity 
approximation. An open problem is that the dilaton remains 
undetermined, consistently with the perturbative nature of the 
mechanism.
To fix the dilaton, one possibility is obviously to turn on at the 
same time closed string 3-form fluxes, or alternatively to use 
non-perturbative effects, such as multiple gaugino condensation in 
the unbroken gauge group sector~\cite{Krasnikov:1987jj}.

A different direction is to study moduli stabilization in vacua with 
broken supersymmetry using a straightforward extension of the 
mechanism we described in this work. It is also interesting to apply 
this approach in semi-realistic string models using explicit 
constructions of intersecting $D$-brane configurations, that are 
related by T-duality to compactifications in the presence of internal 
magnetic fields.

\section*{Acknowledgements}

We would like to thank C. Angelantonj, R. Blumenhagen, A. Ceresole, M. Grana, 
M. Herbst, P. Kaste, T. Matik, K. Rulik, M. Tuckmantel and A. Uranga 
for useful discussions.
This work was supported in part by the European Commission under the 
RTN contract MRTN-CT-2004-503369, and in part by the 
INTAS contract 03-51-6346.


\appendix
\section{Example of consistent moduli stabilization}\label{app:ex}

In this appendix, we give a numerical example of $D9$ brane
magnetic flux configuration allowing for a consistent stabilization 
of the complex structure and K\"ahler class moduli of the
toroidal $T^6$ type I string compactification.
The fluxes to which the stacks $\sharp 7$ to $\sharp9$ couple are
\bea
(m^7,n^7) & = & \{(-2,1),(0,1),(1,1)\}\nonumber\\
(m^8,n^8) & = & \{(-1,1),(1,1),(-1,1)\}\nonumber\\
(m^9,n^9) & = & \{(0,1),(-1,1),(2,1)\}\nonumber
\label{fluxes_appendix}
\eea 
From equation (\ref{linear_sys}), we get
\be
\left(
\begin{array}{ccc}
-2& 0& 1\\
-1& 1&-1\\
0 &-1& 2
\end{array}
\right)
\left(
\begin{array}{c}
J_2J_3\\
J_{1}J_{3}\\
J_{1}J_{2}
\end{array}
\right)
=(4\pi^2\alpha')^2\left(
\begin{array}{c}
    0\\
    1\\
    0\\
\end{array}
\right)\, .
\label{linear_sys_appendix}
\ee
Thus, our choice of fluxes fixes the K\"ahler 
moduli $(J_{1},J_{2},J_{3})$ to the values:
\be
(J_{1},J_{2},J_{3}) = 4\pi^2\alpha' 
\left(2\sqrt{2},\frac{\sqrt{2}}{2},\sqrt{2}\right)\, .
\label{numJ}
\ee
Moreover, if these stacks contain $N_7=N_8=2$ and $N_9=1$ 
branes, the induced 5-brane tadpoles in the 2-cycles $[x_i,y_i]$ are:
\be
(q_{[x_1,y_1]},q_{[x_2,y_2]},q_{[x_3,y_3]}) =(-4,-2,-2)\, .
\label{charge_appendix}
\ee

Following the tadpole cancellation condition (\ref{tadpole_5}), 
the above R-R charges have to be cancelled by the charges 
induced by the first six stacks of branes. From the positivity 
condition (\ref{tadpole_5_1}) and our orientation choice
(\ref{orientation}), all signs ${\mathcal K}_{rst}$ (\ref{Ksign})
involved in the fluxes of the first six brane stacks are
negative, and thus the winding numbers $n^a_r$, for 
$a=1,\dots,6$, should be chosen negative, as well.
Here, we consider the following choice for the
corresponding fluxes:
\bea
\{(m^1_{x_1y_2},n^1_{x_1y_2}),(m^1_{x_2y_1},n^1_{x_2y_1}),(m^1_{x_3y_3},n^1_{x_3y_3})\} 
& = & \{(1,-1),(1,-1),(0,-1)\}\nonumber\\
\{(m^2_{x_1y_3},n^2_{x_1y_3}),(m^2_{x_3y_1},n^2_{x_3y_1}),(m^2_{x_2y_2},n^2_{x_2y_2})\} 
& = & \{(1,-1),(1,-1),(0,-1)\}\nonumber\\
\{(m^3_{x_1x_2},n^3_{x_1x_2}),(m^3_{y_1y_2},n^3_{y_1y_2}),(m^3_{x_3y_3},n^1_{x_3y_3})\} 
& = & \{(1,-1),(1,-1),(0,-1)\}\nonumber\\
\{(m^4_{x_2x_3},n^4_{x_2x_3}),(m^4_{y_2y_3},n^4_{y_2y_3}),(m^4_{x_1y_1},n^4_{x_1y_1})\} 
& = & \{(1,-1),(1,-1),(0,-1)\}\nonumber\\
\{(m^5_{x_1x_3},n^5_{x_1x_3}),(m^5_{y_1y_3},n^5_{y_1y_3}),(m^5_{x_2y_2},n^5_{x_2y_2})\} 
& = & \{(1,-1),(1,-1),(0,-1)\}\nonumber\\
\{(m^6_{x_2y_3},n^6_{x_2y_3}),(m^6_{x_3y_2},n^6_{x_3y_2}),(m^6_{x_1y_1},n^6_{x_1y_1})\} 
& = & \{(1,-1),(1,-1),(0,-1)\}
\nonumber
\label{fluxes_1-6}
\eea
This choice satisfies the consistency conditions (\ref{restriction_complex}),
(\ref{restriction_complex2}) and (\ref{restriction_k}), as all $k_a =1$ for 
$a=1,\dots,6$. The complex structure is then fixed to the
imaginary diagonal matrix (\ref{complex_fixed1}) with
\be
(\tau_{11},\tau_{22},\tau_{33})=i(1,1,1)\, .
\ee

We have therefore found a $D9$ brane magnetic background 
which fixes the geometry of the torus $T^6$ to be a product
of three square torii with radii $R_{2i-1} = R_{2i}$, $i=1,2,3$,
given by (\ref{numJ}). Let us now choose the number of branes in 
each of the first six stacks in order to cancel the charges 
induced by the last three stacks 7, 8 and 9. We take
$N_4 = N_6 =2$, and $N_1=N_2=N_3=N_5=1$, so that the
induced 5-brane charges (\ref{tadpole_cancellation})
on the three 2-cycles are: 
\be
(q_{[x_1,y_1]},q_{[x_2,y_2]},q_{[x_3,y_3]}) = (4,2,2)\, ,
\label{charge_appendix2}
\ee 
which cancel exactly the charges (\ref{charge_appendix}).
On the other hand, the contribution to the $D9$ brane tadpole is:
\be
\sum_{a=1}^9 \, N_a\, \sum_{rst}\mathcal{K}_{rst}n_{a}^{r}n_{a}^{s}n_{a}^{t} = 13\, ,
\ee
implying that we have to add 3 additional non-magnetized $D9$ 
branes (and their images), in order to satisfy eq.~(\ref{tadpole_9}). 

\section{Moduli stabilizaton with 2 large dimensions}\label{app:led}

We want now to find an appropriate rescaling of the solution in 
the example shown in the previous appendix in order to induce a 
large volume hierarchy (compared to $\alpha'$). Following the
reasoning of section~\ref{sec:led}, as an illustration, 
we are interested in triggering two large dimensions. To do that, 
we rescale the component $[x_3,y_3]$ of the fluxes 
(\ref{fluxes_appendix}) by a factor $\Lambda=N^2$, for some
integer $N$. The system of conditions on the K\"ahler 
moduli then reads:
\be
\left(
\begin{array}{ccc}
-2& 0& N^2\\
-1& 1&-N^2\\
0 &-1& 2N^2
\end{array}
\right)
\left(
\begin{array}{c}
J^\Lambda_{2}J^\Lambda_{3}\\
J^\Lambda_{1}J^\Lambda_{3}\\
J^\Lambda_{1}J^\Lambda_{2}
\end{array}
\right)
=(4\pi^2\alpha')^2\left(
\begin{array}{c}
    0\\
    N^2\\
    0\\
\end{array}
\right),
\label{linear_sys_appendixB}
\ee
with solution: 
\be
J^\Lambda = 4\pi^2\alpha' 
\left(2\sqrt{2},\frac{\sqrt{2}}{2},\sqrt{2}N^2 \right)\, .
\ee
The induced 5-brane charges (\ref{q5R}) in the different 2-cycles are 
\be
\left(
    q_{[x_1,y_1]},
    q_{[x_2,y_2]},
    q_{[x_3,y_3]}
\right)
=
\left(
 -4N^2,
 -2N^2,
 -2
\right)\, ,
\label{tadpole_sys_appendix}
\ee
for the same brane stack multiplicities we used in Appendix A.

Let us now find a flux configuration in the first six brane stacks in 
a way to cancel these charges, and such that the complex structure 
moduli $\tau$ remain at the value:
\be
(\tau_{11},\tau_{22},\tau_{33})=i(1,1,1)\, .
\ee
This can be achieved by choosing the ratios of fluxes equal to one, 
$k_a = 1$,  $a=1,\dots,6$. These stacks induce then six dimensional 
charges given by:
\be
\sum_{a=1}^6\left(
\begin{array}{c}
   q_{[x_1,y_1]}^{a} \\
   q_{[x_2,y_2]}^{a} \\
   q_{[x_3,y_3]}^{a} \\
\end{array}
\right)
=\left(
\begin{array}{c}
  2 (m^{4,\Lambda}_{x_2x_3})^2   +  2 (m^{6,\Lambda}_{x_2y_3})^2\\
    (m^{2,\Lambda}_{x_1y_3})^2   +    (m^{5,\Lambda}_{x_1x_3})^2\\
    (m^{1,\Lambda}_{x_1y_2})^2   +    (m^{3,\Lambda}_{x_1x_2})^2\\
\end{array}
\right).
\ee
They are all positive and can cancel the tadpoles of 
(\ref{tadpole_sys_appendix}) by choosing:
\be
 m^{4,\Lambda}_{x_2x_3} = m^{6,\Lambda}_{x_2y_3} = m^{5,\Lambda}_{x_1x_3} = m^{2,\Lambda}_{x_1y_3} = N\, .
\ee
With this choice of configuration, the geometry of the internal space 
is fixed to a product of three square torii with radii $R_{2i-1}=R_{2i}$,
and
\be
R_{1}^2 = 2\sqrt{2}\alpha' \,\,\, , \,\,\, 
R_{3}^2 = \frac{\sqrt{2}}{2}\alpha' \,\,\,\, ,\,\,\, 
R_{5}^2 = \sqrt{2} \, N^2 \,\alpha'\, .
\ee


\end{document}